\documentclass[letterpaper, 10 pt, conference]{ieeeconf}  

\IEEEoverridecommandlockouts                              

\title{Adaptive Game-Theoretic Decision Making \\ for Autonomous Vehicle Control at Roundabouts}

\author{Ran Tian, Sisi Li, Nan Li, Ilya Kolmanovsky, Anouck Girard and Yildiray Yildiz 
\thanks{This research has been supported by the National Science Foundation Award Number CNS 1544844.}
\thanks{Ran Tian and Sisi Li are with the Robotics Institute, University of Michigan, Ann Arbor, MI 48109, USA
        {\tt\small \{tianran,sisli\}@umich.edu}. 
        Nan Li, Ilya Kolmanovsky, and Anouck Girard are with the Department of Aerospace Engineering,
        University of Michigan, Ann Arbor, MI 48109, USA
        {\tt\small \{nanli,ilya,anouck\}@umich.edu}. 
        Yildiray Yildiz is with the Department of Mechanical Engineering,
        Bilkent University, Ankara, Turkey {\tt\small yyildiz@bilkent.edu.tr}.}
}

\usepackage{float}
\usepackage[caption=false]{subfig}
\usepackage{epsfig}
\usepackage{epstopdf}
\usepackage{bbm}
\usepackage{epsfig} 
\usepackage{amsmath} 
\usepackage{amssymb}  
\usepackage{bm}
\usepackage{epstopdf}
\usepackage{color}
\usepackage[linesnumbered]{algorithm2e}
\DeclareMathOperator*{\argmax}{arg\,max}

\usepackage{comment}
\begin{document}
\maketitle

\begin{abstract}
In this paper, we propose a decision making algorithm for autonomous vehicle control at a roundabout intersection. The algorithm is based on a game-theoretic model representing the interactions between the ego vehicle and an opponent vehicle, and adapts to an online estimated driver type of the opponent vehicle. Simulation results are reported.
\end{abstract}



%
\IEEEpeerreviewmaketitle

\section{Introduction}

Autonomous vehicle control in urban traffic is still facing enormous challenges. Many of these challenges involve intersection traffic scenarios. According to \cite{safety_intersection}, almost $40\%$ of car crashes in the U.S. are intersection related. In an intersection scenario, it is typical to have multiple traffic participants interacting with each other. A driver or an automation controlling a vehicle at an intersection should account for these interactions in her/its decision making.

Intersections are usually categorized into two types: signalized intersections and unsignalized intersections, of which roundabout is a particular kind \cite{safety_roundabout}. At a signalized intersection, the motions of all vehicles and of all other traffic participants are guided by traffic lights or traffic signs, which act as centralized traffic controls. On the other hand, at an unsignalized intersection, the driver/automation controlling a vehicle needs to decide on her/its own, whether, when and how to enter and cross the intersection, in which accounting for the interactions between traffic participants is particularly important, as each participant's actions influence and are also influenced by the actions of the other participants.

Game theory is a useful tool to model the interactions between strategic decision makers. Game-theoretic modelings of driver and vehicle interactions in highway traffic scenarios for use in autonomous vehicle control development are discussed in \cite{li7993050, li2019game, talebpour2015modeling}. Game-theoretic autonomous vehicle control algorithms for intersection traffic scenarios are proposed in \cite{Nanintersection, 7313157}. In \cite{Nanintersection}, vehicle-to-vehicle interactions at an unsignalized four-way intersection are modeled using a game-theoretic framework, and then an autonomous vehicle controller for such an intersection is developed based on the vehicle interaction model.

A roundabout intersection involves a merge/de-merge type traffic where a vehicle first merges into the center circle, travels counter-clockwise (in the U.S.), and then de-merges from the circle \cite{roundabout}. Autonomous vehicle control particularly for roundabouts have been investigated in \cite{perez2011autonomous, rastelli2015fuzzy}, where vehicle interactions in multi-vehicle scenarios are not considered.

In this paper, we exploit the game-theoretic vehicle interaction modeling framework in \cite{Nanintersection} to develop an autonomous vehicle controller for a roundabout intersection. The contributions of this paper are: 1) We develop a game-theoretic model representing the interactions between two vehicles at a roundabout intersection. Such a model can have multiple uses, such as for developing autonomous vehicle control systems \cite{Nanintersection} or for the verification, validation, and calibration of such systems \cite{li7993050,li2019game}. We focus on the former in this paper. 2) We propose a decision making algorithm for an ego vehicle at such a roundabout intersection that is based on the vehicle interaction model and adapts to an estimated driver type of an opponent vehicle. 3) We describe an explicit online implementation scheme exploiting function approximation techniques to avoid the need for solving optimization problems related to the algorithm in real time.


\section{Vehicle Model}
\label{sec: vehicle model}
\subsection{Vehicle kinematics}
\label{sec: vehicle kinematics}

The control of a vehicle is typically modeled using a hierarchical architecture \cite{li2017tree}: a high-level decision making layer plans the desired path for the vehicle, and then a low-level dynamics and actuation control layer controls the subsystems, e.g., engine, transmission, steering, etc., to track the references generated by the high level.


In this paper, we focus on the high-level decision making and we use a discrete-time model to represent the vehicle kinematics at a roundabout intersection as follows,
\begin{subequations}\label{equ:kinematics}
\begin{gather}
    x(t+1) = x(t) + v(t)\cos{\big(\theta(t)\big)} \, \Delta t,\\
    y(t+1) = y(t) + v(t)\sin{\big(\theta(t)\big)} \, \Delta t,\\
    v(t + 1) = v(t) + a(t) \, \Delta t,\\
    \theta(t+1) = \theta(t) + \omega(t) \, \Delta t,
\end{gather}
\end{subequations}
where $x(t)$ and $y(t)$ represent, respectively, the vehicle's position in the $x$-direction and $y$-direction at the discrete time $t$; $v(t)$, $a(t)$, $\theta(t)$, and $\omega(t)$ are, respectively, the vehicle's speed, acceleration, yaw angle and yaw rate at $t$; $\Delta t$ is the time step size. An action, denoted by $\gamma$, is an input pair $(a, \omega)$ to the model \eqref{equ:kinematics}. In this paper, we assume that the vehicle can choose actions from a finite action set, $\mathbf{\Gamma} = \{\gamma_1, \gamma_2, \dots , \gamma_m\}$, to execute.

\subsection{Scenario of interest}\label{sec: roundabout}

\begin{figure}[ht]
\begin{center}
\begin{picture}(240, 110.0)
\put(0,-4){\epsfig{file=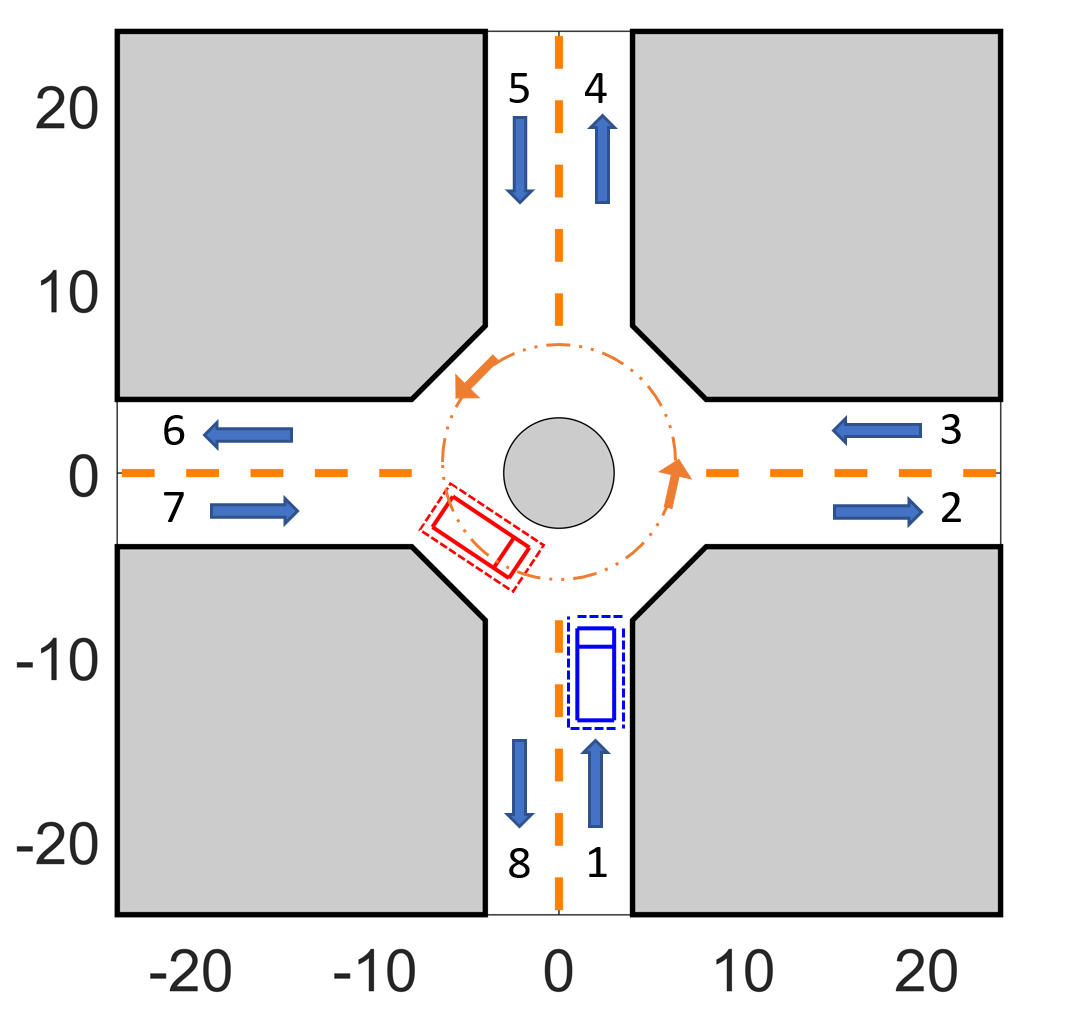, width = 0.465\linewidth}}  
\put(120,-1){\epsfig{file=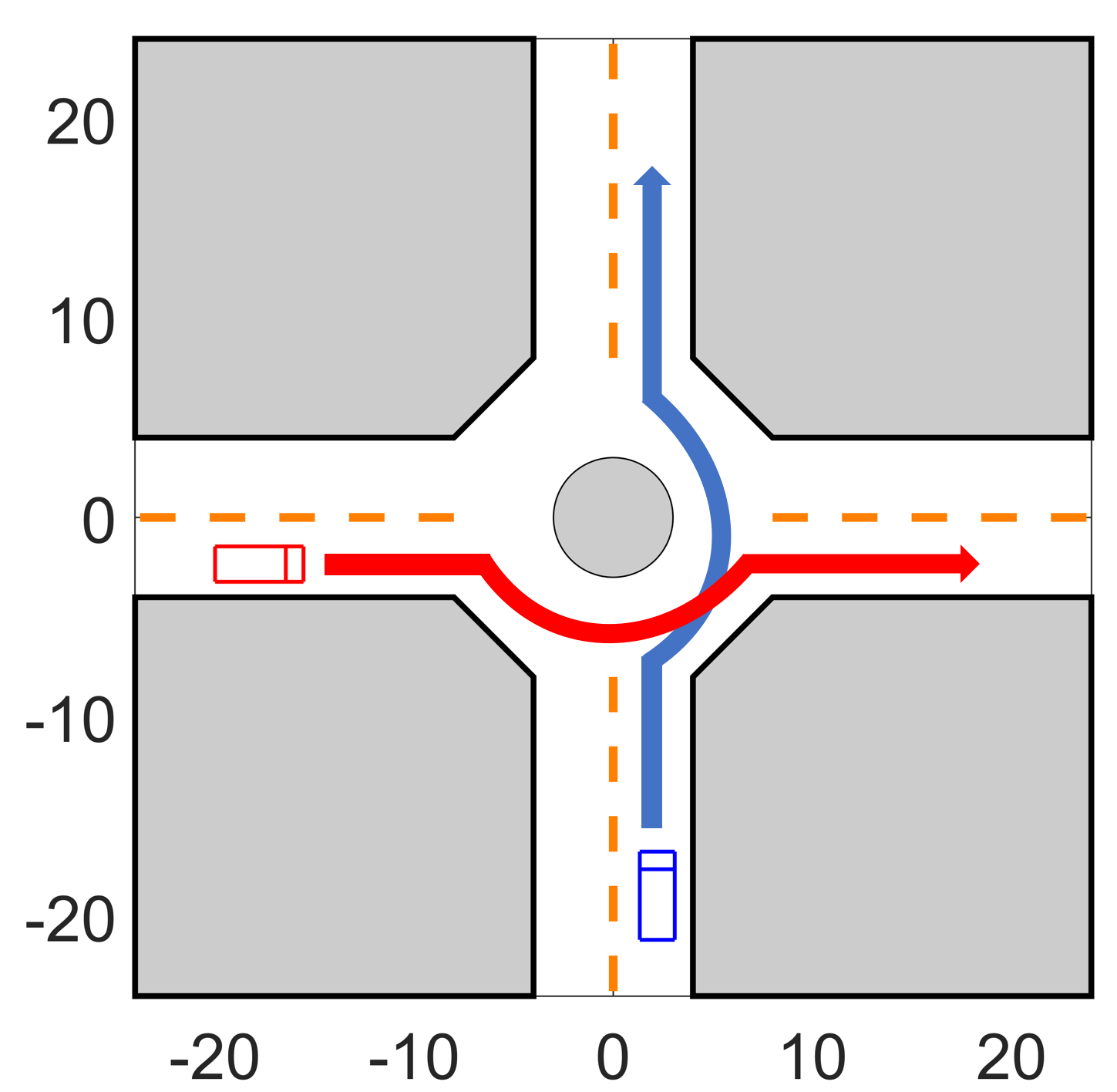, width = 0.445\linewidth}}
\small
\put(95,106){(a)}
\put(215,106){(b)}

\end{picture}
\end{center}
      \caption{The roundabout intersection and traffic scenario to be considered.}
      \label{fig: road}
\end{figure}

In this paper, we model the vehicle-to-vehicle interactions at a roundabout intersection, see Fig.~\ref{fig: road}(a). A vehicle can enter or exit the roundabout in the directions indicated by the blue arrows. When in the roundabout, all vehicles must travel counter-clockwise (indicated by the orange arrows). In this paper, we consider the interactions between two vehicles, the ego vehicle (represented by the blue solid box in Fig.~\ref{fig: road}(a)) and the opponent vehicle (represented by the red solid box in Fig.~\ref{fig: road}(a)). The double lines indicate the front ends of the vehicles. 

\vspace{-0.5cm}
\subsection{Reward function}\label{sec: reward function}

The vehicle decision making process is based on receding-horizon optimal control: At each time step $t$, the vehicle selects a sequence of actions $\Gamma(t) = \{\gamma(t),\cdots,\gamma(t+n-1)\}$ to maximize a cumulative reward over a horizon of length $n$, $\mathcal{R}(t) = \sum\limits_{j=1}^{n} \lambda^{j-1} R(t+j)$, where $R(t+j)$ represents a stage reward at step $t+j$ over the horizon, and $\lambda \in [0,1]$ is a discount factor.

The stage reward function, $R(t)$, is defined as 
\begin{equation}
\label{eq: reward}
    R(t) = \mathbf{w}^\intercal \mathbf{\Phi}(t),
\end{equation}
where $\mathbf{\Phi}(t) = \left[\phi_1(t),\, \phi_2(t),\, \phi_3(t),\, \phi_4(t),\, \phi_5(t),\, \phi_6(t)\right]^\intercal$ 
is the feature vector at $t$, and $\mathbf{w}$ is the weight vector that contains the weights for the features.

The feature $\phi_1 (t)$ is an indicator that characterizes the collision status of the vehicle.  We bound the geometric contour of the vehicle by a rectangle (the dashed boxes in Fig.~\ref{fig: road}(a)). We refer to this rectangle as the collision avoidance zone (c-zone). If the ego vehicle's c-zone has an overlap with the opponent vehicle's c-zone, which indicates a danger of collision, $\phi_1(t) = -1 $; $\phi_1(t) = 0 $ otherwise. In this paper, the size of the c-zone is $ 5 \ \text{[m]} \times 2 \ \text{[m]}$.

The feature $\phi_2(t)$ is an indicator that characterizes the on-road status of the vehicle. If the ego vehicle's c-zone crosses the road boundaries, i.e., the ego vehicle's c-zone has an overlap with the gray areas in Figure~\ref{fig: road}(a), $\phi_2(t) = -1$; $\phi_2(t) = 0$ otherwise.

The feature $\phi_3(t)$ characterizes the distance-to-objective status of the vehicle. We define $\phi_3(t)$ as
\begin{equation}
    \phi_3(t) = -\, |x_r-x(t)| - |y_r-y(t)|,
\end{equation}
where $(x_r,y_r)$ are the coordinates of a reference point on the vehicle's objective lane. 

The feature $\phi_4(t)$ characterizes the safe separation of the ego vehicle from the opponent vehicle. When driving in traffic, a vehicle is supposed to keep a reasonable distance from its surrounding vehicles to improve safety. We define a safety zone (s-zone) that over-bounds the vehicle's c-zone with a safety margin. If the ego vehicle's s-zone has an overlap with the opponent vehicle's s-zone, $\phi_4(t) = -1$; $\phi_4(t) = 0$ otherwise. In this paper, the s-zone is concentric with the c-zone and is $8 \ \text{[m]} \times 2.4 \ \text{[m]}$ in size. 

The feature $\phi_5(t)$ penalizes crossing the lane markings that separate traffic of opposite directions and driving into a wrong lane (not the vehicle's objective lane) when exiting the roundabout. If either occurs, $\phi_5(t) = -1$; $\phi_5(t) = 0$ otherwise.

The feature $\phi_6(t)$ rewards the vehicle's speed, defined as $\phi_6(t) = v(t)$. 


\section{Game Theoretic Decision Making}
\label{sec: level-k}


Based on the definitions of the features, the rewards of a vehicle not only depend on its own states and actions, but also depend on the states and actions of its opponent vehicle (e.g., $\phi_1(t)$ and $\phi_4(t)$). Such an interdependence reflects the interactive nature of vehicle decision making in a multi-vehicle traffic scenario. The receding-horizon optimal control problem is thus formulated as: At each time step $t$, to select
\begin{align}\label{eq: find action}
    & \Gamma_{\text{ego}}^*(t) = \{\gamma_{\text{ego}}^*(t),\cdots,\gamma_{\text{ego}}^*(t+n-1)\} = \\
    & \argmax_{\gamma_{\text{ego}}(t+j) \in \mathbf{\Gamma}} \sum_{j = 0}^{n-1} \lambda^j R \big(\mathbf{X}(t+j), \gamma_{\text{ego}}(t+j), \gamma_{\text{opp.}}(t+j)\big), \nonumber
\end{align}
and execute $\gamma_{\text{ego}}^*(t)$ over one time step, where $\mathbf{X}(t) = [x_{\text{ego}}(t), \ y_{\text{ego}}(t), \ \theta_{\text{ego}}(t), \ v_{\text{ego}}(t), \ x_{\text{opp.}}(t), \ y_{\text{opp.}}(t), \ \theta_{\text{opp.}}(t), \\ v_{\text{opp.}}(t)]^\intercal$ represents the state of the traffic, which contains both the ego vehicle's states and the opponent vehicle's states; $\gamma_{\text{ego}}(t+j)$ is the ego vehicle's action at $t+j$ over the horizon and is to be optimized, and $\gamma_{\text{opp.}}(t+j)$ is the opponent vehicle's action at $t+j$ over the horizon and is to be predicted. We note that for the two interacting vehicles, either is the ``ego vehicle'' from its own perspective, and is also the ``opponent vehicle'' from the other's perspective, that is, \eqref{eq: find action} can be used to describe the decision making of either of the two vehicles.

To obtain $\Gamma_{\text{ego}}^*(t)$, the ego vehicle needs to predict the sequence $\Gamma_{\text{opp.}}(t) = \{\gamma_{\text{opp.}}(t),\cdots,\gamma_{\text{opp.}}(t+n-1)\}$, which contains the actions that the opponent vehicle executes over the prediction horizon. Approaches such as data-driven driver behavior prediction may be used for such predictions \cite{8290702}. In this paper, we exploit a game-theoretic approach for such predictions as it accounts for vehicle-to-vehicle interactions. 

Our approach is based on the level-$k$ game theory \cite{costa2009comparing}. Level-$k$ game theory relies on a hierarchical cognitive structure to model human reasoning in games. Each player is assumed to be of a particular reasoning level, indicated by $k \in \{0,1,\cdots\}$. A level-$k$ player assumes that all other players can be modeled as level-($k-1$) reasoners and acts accordingly. The development of the reasoning hierarchy starts from a level-$0$ player's model, which usually represents instinctive decision making without accounting for the interactions between players. Then a level-$1$ player's model can be obtained by assuming that all the players except for the ego player can be modeled as level-$0$ players. Based on this assumption, a level-$1$ player's model can predict the actions of the other level-$0$ players, e.g., the sequence $\Gamma_{\text{opp.}}(t)$ in \eqref{eq: find action}, and then compute its own optimal actions $\Gamma_{\text{ego}}^*(t)$ based on \eqref{eq: find action}. This procedure continues -- a level-$k$ player's model is generated by assuming that all other players act according to level-($k-1$) models, until the desired highest level is obtained. This reasoning hierarchy has been observed in human interactions for other application domains by experimental studies and it is shown that humans are commonly level-0, 1 and 2 reasoners \cite{costa2009comparing,costa2006cognition}. 

In \cite{Nanintersection}, such a level-$k$ game-theoretic framework is employed to model the driver and vehicle interactive behavior at an unsignalized four-way intersection. By assuming that a level-$0$ driver treats the other vehicles on the road as stationary obstacles (this way, a level-$0$ driver may be viewed as an aggressive driver in human traffic, who usually assumes that the others will yield the right of way) and constructing the corresponding level-$1$ and $2$ driver models using the procedure described above, we observe that the behavior of a level-$0$ driver and that of a level-$2$ driver are similar as they both represent aggressive drivers. Therefore, instead of considering three driver levels (level-$0$, $1$ and $2$), we consider two driver types, type-$1$ and $2$, in this paper.

A type-$1$ driver model represents a conservative driver who predicts the opponent vehicle's actions based on the assumption that the opponent vehicle's driver is an instinctive decision maker, who maximizes her cumulative reward \eqref{eq: find action} by treating the other vehicles on the road as stationary obstacles. On the other hand, a type-$2$ driver model represents an aggressive driver who predicts the opponent vehicle's actions by assuming that the opponent vehicle's driver is a type-$1$ driver. Indeed, the type-$1/2$ driver model meets the level-$1/2$ driver model defined in the level-$k$ game-theoretic framework described above.



Specifically, a type-$1$ driver selects actions, $\Gamma_{\text{ego}}^{(1)}(t) = \{\gamma_{\text{ego}}^{(1)}(t),\cdots,\gamma_{\text{ego}}^{(1)}(t+n-1)\}$, based on
\begin{align}\label{equ:T1_act}
    & \Gamma_{\text{ego}}^{(1)} (t) = \\[-2pt]
    & \argmax_{\gamma_{\text{ego}}(t+j) \in \mathbf{\Gamma}} \sum_{j = 0}^{n-1} \lambda^j R \big(\mathbf{X}(t+j), \gamma_{\text{ego}}(t+j), \gamma^{(0)}_{\text{opp.}}(t+j)\big), \nonumber
\end{align}
where $\gamma^{(0)}_{\text{opp.}}(t+j)$ represents the predicted opponent vehicle's action at $t+j$ over the horizon under the assumption that the opponent vehicle's driver treats the ego vehicle as a stationary obstacle. Note that based on this assumption, the actions $\gamma^{(0)}_{\text{opp.}}(t+j)$, $j=0,\cdots,n-1$, are independent of $\Gamma_{\text{ego}}^{(1)} (t)$, and thus can be determined first. 

Similarly, a type-$2$ driver selects actions, $\Gamma_{\text{ego}}^{(2)}(t)$, based on
\begin{align}\label{equ:T2_act}
    & \Gamma_{\text{ego}}^{(2)} (t) = \\[-2pt]
    & \argmax_{\gamma_{\text{ego}}(t+j) \in \mathbf{\Gamma}} \sum_{j = 0}^{n-1} \lambda^j R \big(\mathbf{X}(t+j), \gamma_{\text{ego}}(t+j), \gamma^{(1)}_{\text{opp.}}(t+j)\big), \nonumber
\end{align}
where $\{\gamma_{\text{opp.}}^{(1)}(t),\cdots,\gamma_{\text{opp.}}^{(1)}(t+n-1)\}$ are computed using \eqref{equ:T1_act} by switching the roles of ``ego'' and ``opp.''


\section{Adaptive Game-Theoretic Autonomous Vehicle Control}
\label{sec: controller}

\subsection{Adaptive control strategy}\label{sec: controllerA}


Based on the type-$1,2$ driver models, we develop an autonomous vehicle (AV) control  algorithm. At each time step $t$, the AV controller predicts the opponent vehicle's actions over the horizon, $\Gamma_{\text{opp.}}(t)$, based on an estimate of the opponent vehicle's driver type. Specifically, $\Gamma_{\text{opp.}}(t) = \Gamma_{\text{opp.}}^{(1)}(t)$ computed using \eqref{equ:T1_act} if the opponent vehicle's driver is estimated as type-$1$, and $\Gamma_{\text{opp.}}(t) = \Gamma_{\text{opp.}}^{(2)}(t)$ computed using \eqref{equ:T2_act} if the opponent vehicle's driver is estimated as type-$2$ (with the roles of ``ego'' and ``opp.'' switched). Then, the controller computes the optimal actions $\Gamma_{\text{ego}}^*(t)$ for the AV to execute using \eqref{eq: find action} with the predicted $\Gamma_{\text{opp.}}(t)$ substituted in.


The estimate of the opponent vehicle's driver type is updated after each time step. Specifically, at time $t+1$, we compare the opponent vehicle's actually applied action $\gamma_{\text{opp.}}(t)$ to the first action $\gamma_{\text{opp.}}^{(1)} (t)$ in the sequence $\Gamma_{\text{opp.}}^{(1)} (t)$ and to the first action $\gamma_{\text{opp.}}^{(2)} (t)$ in the sequence $\Gamma_{\text{opp.}}^{(2)} (t)$. If $\gamma_{\text{opp.}}(t)$ matches $\gamma_{\text{opp.}}^{(\eta)} (t)$ better for some $\eta \in \{1,2\}$, we increase the probability that the opponent vehicle's driver can be modeled as type-$\eta$.

Such an adaptive game-theoretic AV control strategy has been tested versus different driver models for the opponent vehicle at a non-roundabout type four-way intersection in \cite{Nanintersection}, and exhibits reasonable performance. The action sequences $\Gamma_{\text{opp.}}^{(1)}(t)$, $\Gamma_{\text{opp.}}^{(2)}(t)$, and $\Gamma_{\text{ego}}^*(t)$ are computed based on \eqref{equ:T1_act}, \eqref{equ:T2_act}, and \eqref{eq: find action} using a tree-search based method, which can be computationally demanding when the prediction horizon is large. Thus, in what follows, we propose an explicit online implementation scheme exploiting function approximation techniques, to avoid the need for solving optimization problems in real time.

\subsection{Explicit online implementation}

We note that a function $g: \big(\mathbf{X}(t),\eta(t)\big) \mapsto \gamma_{\text{opp.}}^{(\eta(t))} (t)$, where $\eta(t) \in \{1,2\}$ indicates the opponent vehicle's driver type at $t$, is implicitly defined by \eqref{equ:T1_act} and \eqref{equ:T2_act} (with the roles of ``ego'' and ``opp.'' switched). Given the traffic state $\mathbf{X}(t)$, we denote the $\mathbf{X}(t)$-section of $g$ by $g^{\mathbf{X}(t)}: \eta(t) \mapsto \gamma_{\text{opp.}}^{(\eta(t))} (t)$. The procedure to update the opponent vehicle's driver type estimate described in Section~\ref{sec: controllerA} requires the inverse map of $g^{\mathbf{X}(t)}$, $\big(g^{\mathbf{X}(t)}\big)^{-1}: \gamma_{\text{opp.}}^{(\eta(t))} (t) \mapsto \eta(t) \in \{1,2\}$. Such an inverse map $\big(g^{\mathbf{X}(t)}\big)^{-1}$ may be multi-valued because at some traffic states, the opponent vehicle may select the same action $\gamma_{\text{opp.}}(t)$ regardless of its driver types, e.g., when the two vehicles are far from each other. To have a single-valued function so that it can be fitted using a function approximator, we want to restrict $\big(g^{\mathbf{X}(t)}\big)^{-1}$ to be defined on a subset of the state space of $\mathbf{X}(t)$, denoted by $\mathcal{X}_{\text{critical}}$, such that $\big(g^{\mathbf{X}(t)}\big)^{-1}$ is single-valued on $\mathcal{X}_{\text{critical}}$. To find the set $\mathcal{X}_{\text{critical}}$ is to find the subset of the state space of $\mathbf{X}(t)$ where the action $\gamma_{\text{opp.}}^{(1)} (t)$ selected by a type-1 driver model is different from the action $\gamma_{\text{opp.}}^{(2)} (t)$ selected by a type-2 driver model. More specifically, for each $\mathbf{X}(t)$, we compare $\gamma_{\text{opp.}}^{(1)} (t)$ and $\gamma_{\text{opp.}}^{(2)} (t)$: if $\gamma_{\text{opp.}}^{(1)} (t) \neq \gamma_{\text{opp.}}^{(2)} (t)$, $\mathbf{X}(t) \in \mathcal{X}_{\text{critical}}$; $\mathbf{X}(t) \notin \mathcal{X}_{\text{critical}}$ otherwise. The construction of $\mathcal{X}_{\text{critical}}$ can be done offline, then approximated and described using a clustering neural network (NN$_B$).

After the set $\mathcal{X}_{\text{critical}}$ is identified, when $\mathbf{X}(t) \in \mathcal{X}_{\text{critical}}$, through the function $\big(g^{\mathbf{X}(t)}\big)^{-1} \big(\gamma_{\text{opp.}}(t)\big)$, we obtain an estimate of the opponent vehicle's driver type at $t$, $\eta(t)$ (assuming that $\gamma_{\text{opp.}}(t) = \gamma_{\text{opp.}}^{(\eta)} (t)$, for some $\eta \in \{1,2\}$). The function $g^{-1}\big|_{\mathcal{X}_{\text{critical}}} : \big(\mathbf{X}(t),\gamma_{\text{opp.}}(t)\big) \mapsto \eta(t) \in \{1,2\}$ is constructed offline and approximated using a neural network (NN$_C$).

Taking historical data into account, we update the AV controller's belief on the opponent vehicle's driver type based on
\begin{equation}\label{equ:update}
\mathbb{P}^{(2)}(t+1) = (1-\beta)\mathbb{P}^{(2)}(t) + \beta \, \mathbb{I}\{\eta(t)=2\},    
\end{equation}
where $\mathbb{P}^{(2)}(t)$ represents the probability that the opponent vehicle's driver can be modeled as type-$2$, $\beta \in [0,1]$ is an estimate update step size, and $\mathbb{I}\{\eta(t)=2\}$ is an indicator function, taking $1$ if the event $\{\eta(t) = 2\}$ is true and $0$ otherwise.

Based on $\mathbb{P}^{(2)}(t)$, the AV controller predicts the opponent vehicle's actions over the horizon, $\Gamma_{\text{opp.}}(t)$, using different models: If $\mathbb{P}^{(2)}(t) < 0.5$, the opponent vehicle's driver is more likely to be a type-$1$ driver, and thus the AV controller sets $\Gamma_{\text{opp.}}(t) = \Gamma_{\text{opp.}}^{(1)}(t)$. Otherwise, the opponent vehicle's driver is more likely to be a type-$2$ driver, and thus the AV controller sets $\Gamma_{\text{opp.}}(t) = \Gamma_{\text{opp.}}^{(2)}(t)$. Finally, the AV controller computes the optimal actions $\Gamma_{\text{ego}}^*(t)$ using \eqref{eq: find action} with the predicted $\Gamma_{\text{opp.}}(t)$ substituted in. Similar to the function $g$ implicitly defined by \eqref{equ:T1_act} and \eqref{equ:T2_act}, a control policy $\pi_{\text{ego}}: \big(\mathbf{X}(t),\eta(t)\big) \mapsto \gamma_{\text{ego}}^*(t) \in \mathbf{\Gamma}$, where $\eta(t) \in \{1,2\}$ indicates the estimated opponent vehicle's driver type at $t$, is implicitly defined by \eqref{eq: find action}. The policy $\pi_{\text{ego}}$ is constructed offline and approximated using a neural network (NN$_A$).

\begin{figure}[ht]
\centering
\includegraphics[width=0.44\textwidth]{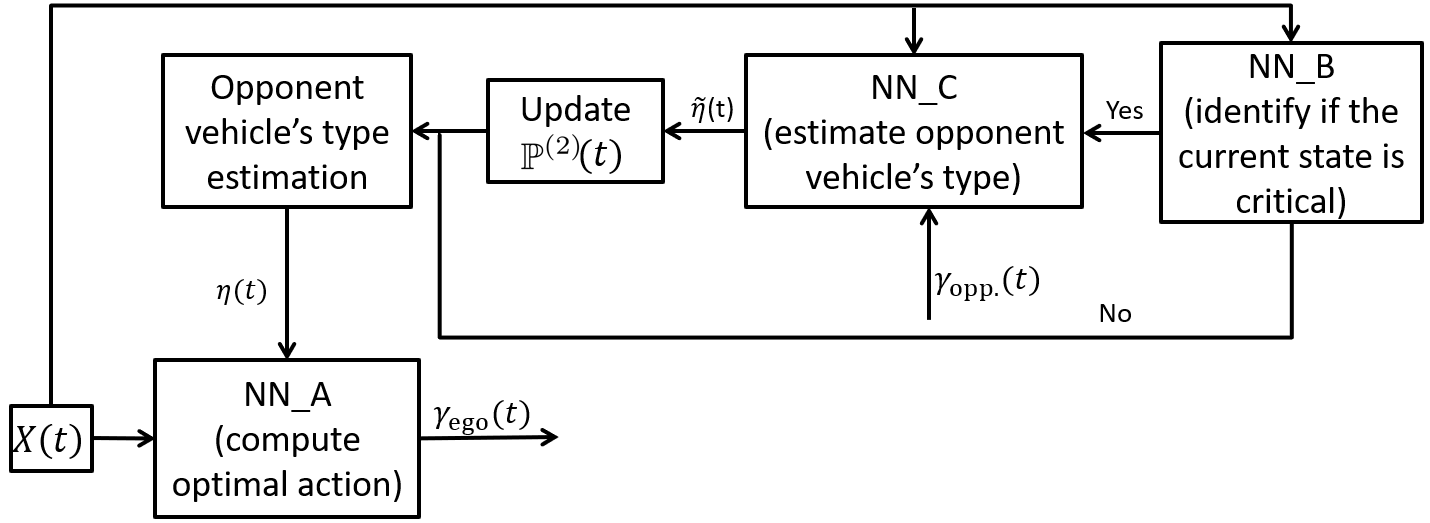}
\caption{Structure of the proposed autonomous vehicle controller.}
\label{fig: NN_diagram}
\end{figure}

With the use of the above three neural networks, we move the computations to solve the optimization problems \eqref{eq: find action}, \eqref{equ:T1_act}, and \eqref{equ:T2_act} from online to offline. The online control is policy-based and requires only neural network evaluations, where the policy adapts to the opponent vehicle's driver by the adaptation law \eqref{equ:update}. The overall structure of the AV controller is shown in Fig~\ref{fig: NN_diagram}.

\subsection{Controller training}
\label{sec: controller training}


We randomly create states $\mathbf{X}(t)$, and compute $\Gamma_{\text{opp.}}^{(1)}(t)$ using \eqref{equ:T1_act} and $\Gamma_{\text{opp.}}^{(2)}(t)$ using \eqref{equ:T2_act} (with the roles of ``ego'' and ``opp.'' switched). If $\gamma_{\text{opp.}}^{(1)} (t) \neq \gamma_{\text{opp.}}^{(2)}(t)$, we label $\mathbf{X}(t)$ by $\zeta\big( \mathbf{X}(t)\big) = 1$; and $\zeta\big(\mathbf{X}(t)\big) = 0 $ otherwise. This labeled data set is used to train NN$_B$.

If $\zeta\big(\mathbf{X}(t)\big) = 1$, we label the pair $\big(\mathbf{X}(t),\gamma_{\text{opp.}}^{(1)}(t)\big)$ by $\eta\big(\mathbf{X}(t),\gamma_{\text{opp.}}^{(1)}(t)\big) = 1$, and label the pair $\big(\mathbf{X}(t),\gamma_{\text{opp.}}^{(2)}(t)\big)$ by $\eta\big(\mathbf{X}(t),\gamma_{\text{opp.}}^{(2)}(t)\big) = 2$. This labeled data set is used to train NN$_C$.

Finally, we compute the optimal actions $\Gamma_{\text{ego}}^*(t,\eta(t)) = \{\gamma_{\text{ego}}^*(t,\eta(t)),\cdots,\gamma_{\text{ego}}^*(t+n-1,\eta(t))\}$ based on \eqref{eq: find action} with $\Gamma_{\text{opp.}}(t) = \Gamma_{\text{opp.}}^{(\eta(t))}(t)$, $\eta(t) = 1,2$, substituted in, and label the pair $\big(\mathbf{X}(t),\eta(t)\big)$ by
$\gamma\big(\mathbf{X}(t),\eta(t)\big) = \gamma_{\text{ego}}^*(t,\eta(t))$. This labeled data set is used to train NN$_A$. 

Table~\ref{table: NN_train} lists the features and labels for the three neural networks. Each data set is split into a training set and a validation set with a ratio of $8$:$2$. We use the same architecture for each neural network (2 convolutional layers followed by 6 fully connected layers) with different hyperparameters. The validation accuracy of each neural network is shown in Table~\ref{table: NN_train}.


\begin{table}[ht]
\centering
\begin{tabular}{ c | c | c| c}
 \hline
 Neural Net & Features & Label & Accuracy \\
 \hline
 NN$_A$ & $\mathbf{X}(t), \eta(t)$& $\gamma_{\text{ego}}^*(t,\eta(t))$ & 97.4\% \\
 \hline
 NN$_B$ &  $\mathbf{X}(t)$ & $\zeta\big(\mathbf{X}(t)\big)$  & 98.2\% \\
 \hline
 NN$_C$ & $\mathbf{X}(t),\gamma_{\text{opp.}}^{(\eta)}(t)$ & $\eta\big(\mathbf{X}(t),\gamma_{\text{opp.}}^{(\eta)}(t)\big)$ & 96.7\% \\
 \hline
\end{tabular}
\caption{Training features, labels and validation accuracy.}
\label{table: NN_train}
\end{table}


\section{Simulation Results}
\label{sec: online simulation}

In this section, we present simulation results to show the performance of the adaptive game-theoretic AV controller. The traffic scenario to be considered is shown in Fig.~\ref{fig: road}(b). The vehicles are assumed to select actions from the set in Table~\ref{table: action}.
The weight in \eqref{eq: reward} is chosen as $\mathbf{w} = [1000,\, 500,\, 5,\, 100,\, 50,\, 1]^\intercal$. The discount factor in the cumulative reward function is $\lambda = 0.8$, and the update step size in the adaptation law \eqref{equ:update} is $\beta = 0.6$.


\begin{table}[h!]
\centering
\begin{tabular}{ c  c  c}
 \hline
action $\gamma$ & $a \ \text{[m}^2/\text{s}]$ & $\omega \ \text{[rad/s]}$  \\
 \hline
 maintain ($\gamma_1$) & 0 & 0\\
 \hline
 accelerate ($\gamma_2$) & 2.5 & 0\\
 \hline
 decelerate ($\gamma_3$) & -2.5 & 0\\
 \hline
 hard brake ($\gamma_4$) & -5 & 0\\
 \hline
 turn left ($\gamma_5$) & 0 & $\pi/4$\\
 \hline
 turn right ($\gamma_6$) & 0 & $-\pi/4$\\
 \hline
\end{tabular}
\caption{Action set $\mathbf{\Gamma}$.}
\label{table: action}
\end{table}

We first test the AV controller's performance versus opponent vehicles controlled by type-1 or 2 drivers. We then test the controller's performance versus driver models that may not act exactly as type-1 or 2 models. In particular, we let human operators control the opponent vehicle using a keyboard. 


\subsection{AV controller versus type-1/2 drivers}

\begin{figure}[ht]
\begin{center}
\begin{picture}(215.0, 370.0)
\put(  10,  282){\epsfig{file=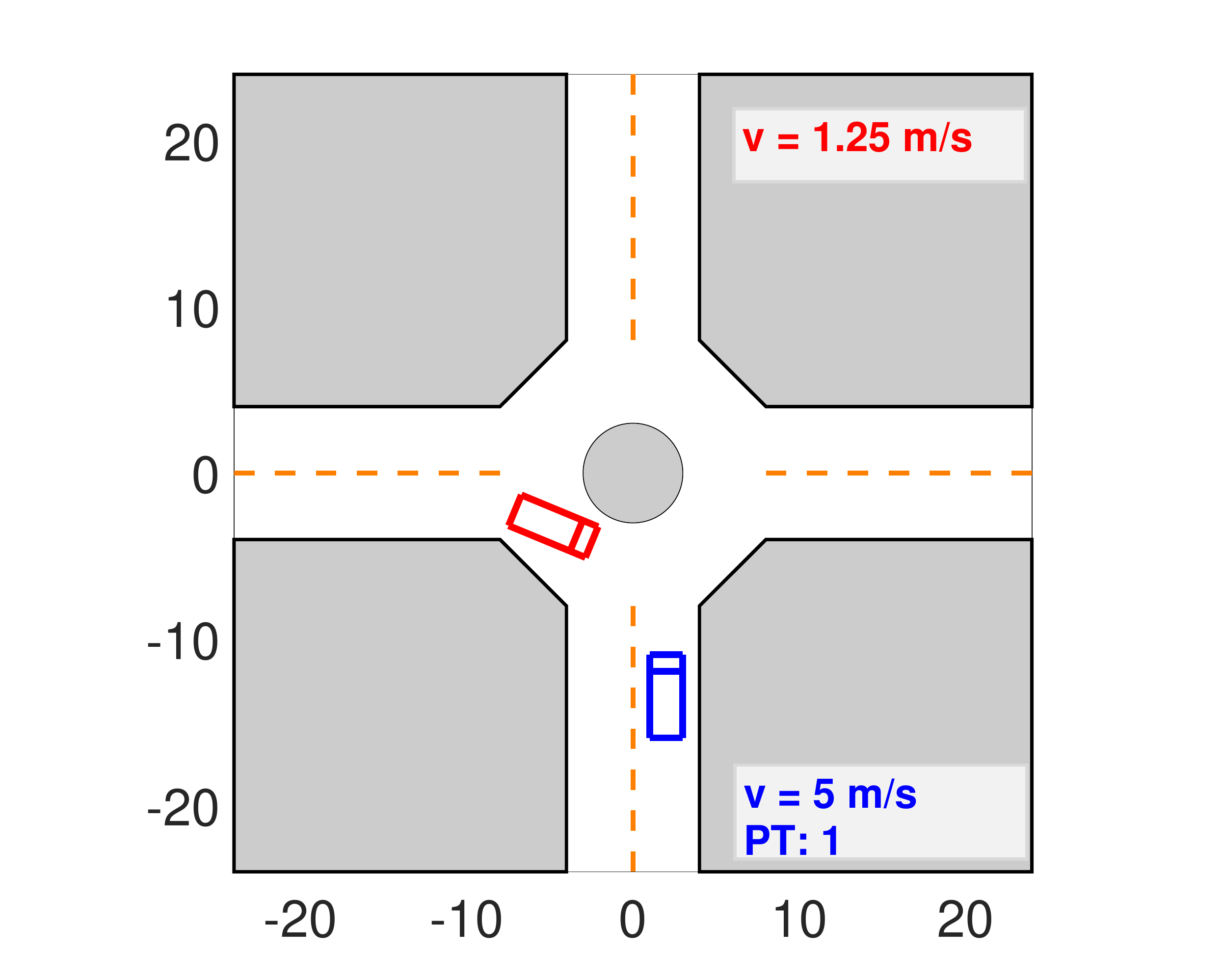,width = 0.4 \linewidth, trim=4.6cm 2cm 0cm 0cm,clip}} 
\put(  120,  282){\epsfig{file=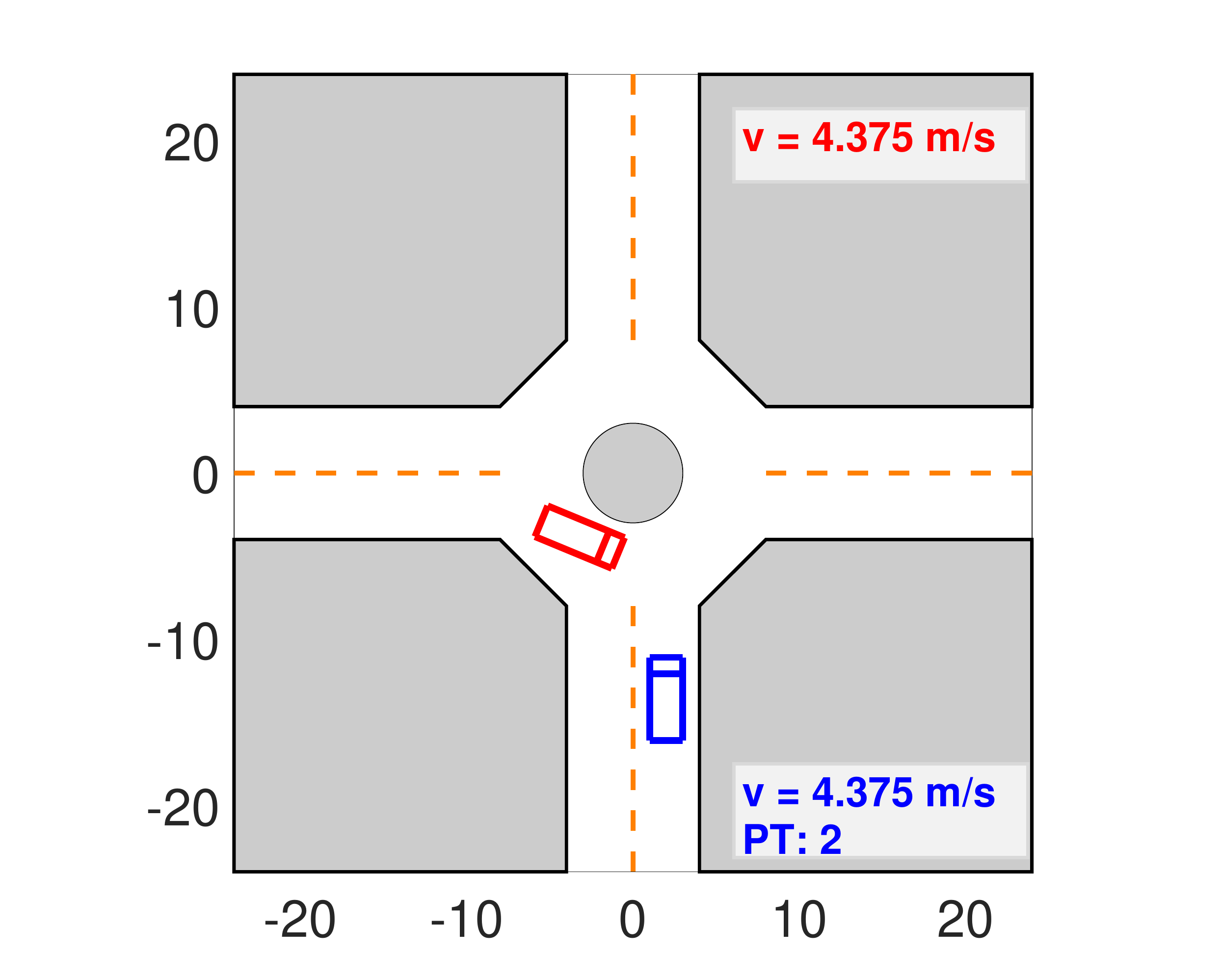,width = 0.4 \linewidth, trim=4.6cm 2cm 0cm 0cm,clip}}  
\put(  10,  188){\epsfig{file=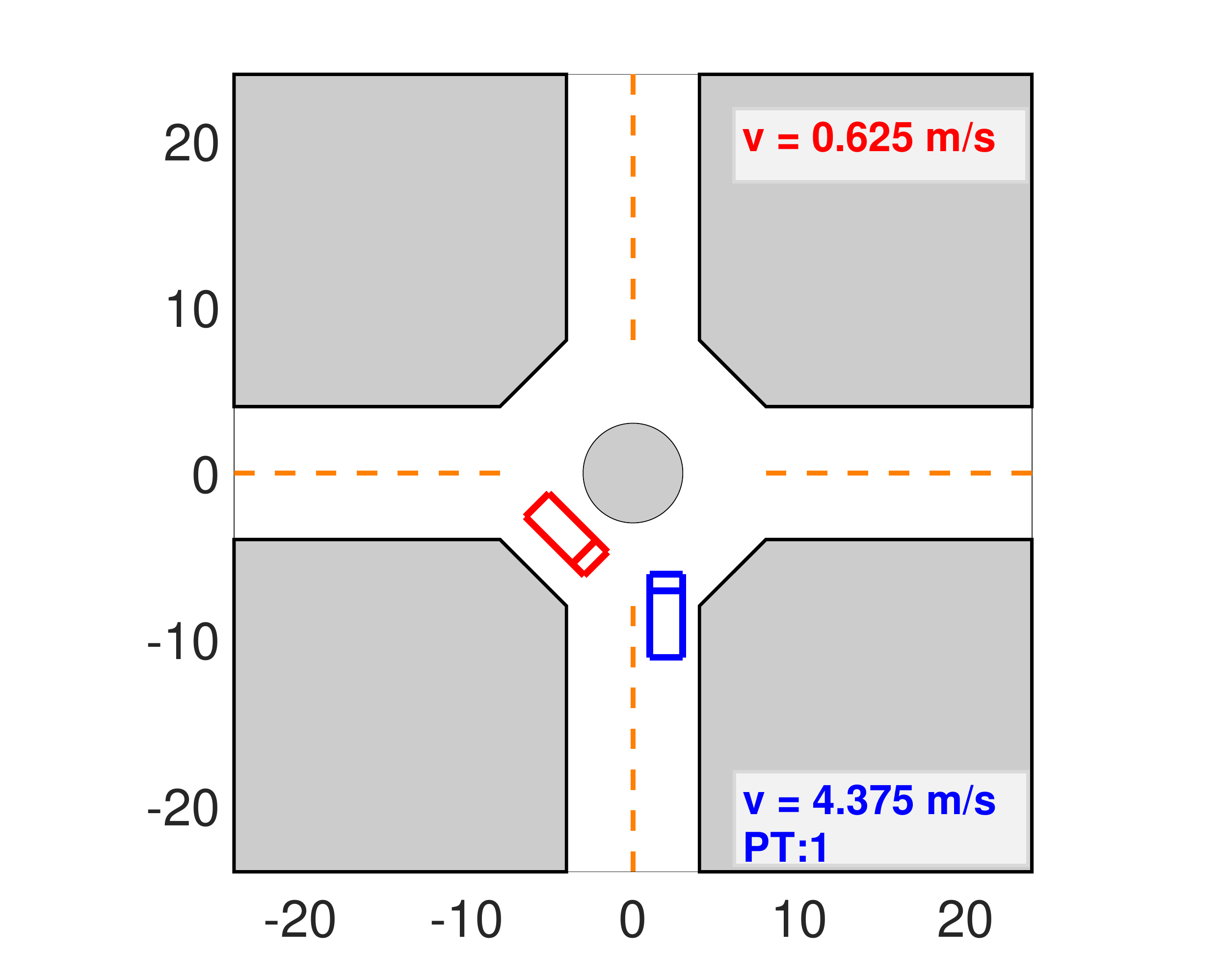,width = 0.4 \linewidth, trim=4.6cm 2cm 0cm 0cm,clip}}  
\put(  120,  188){\epsfig{file=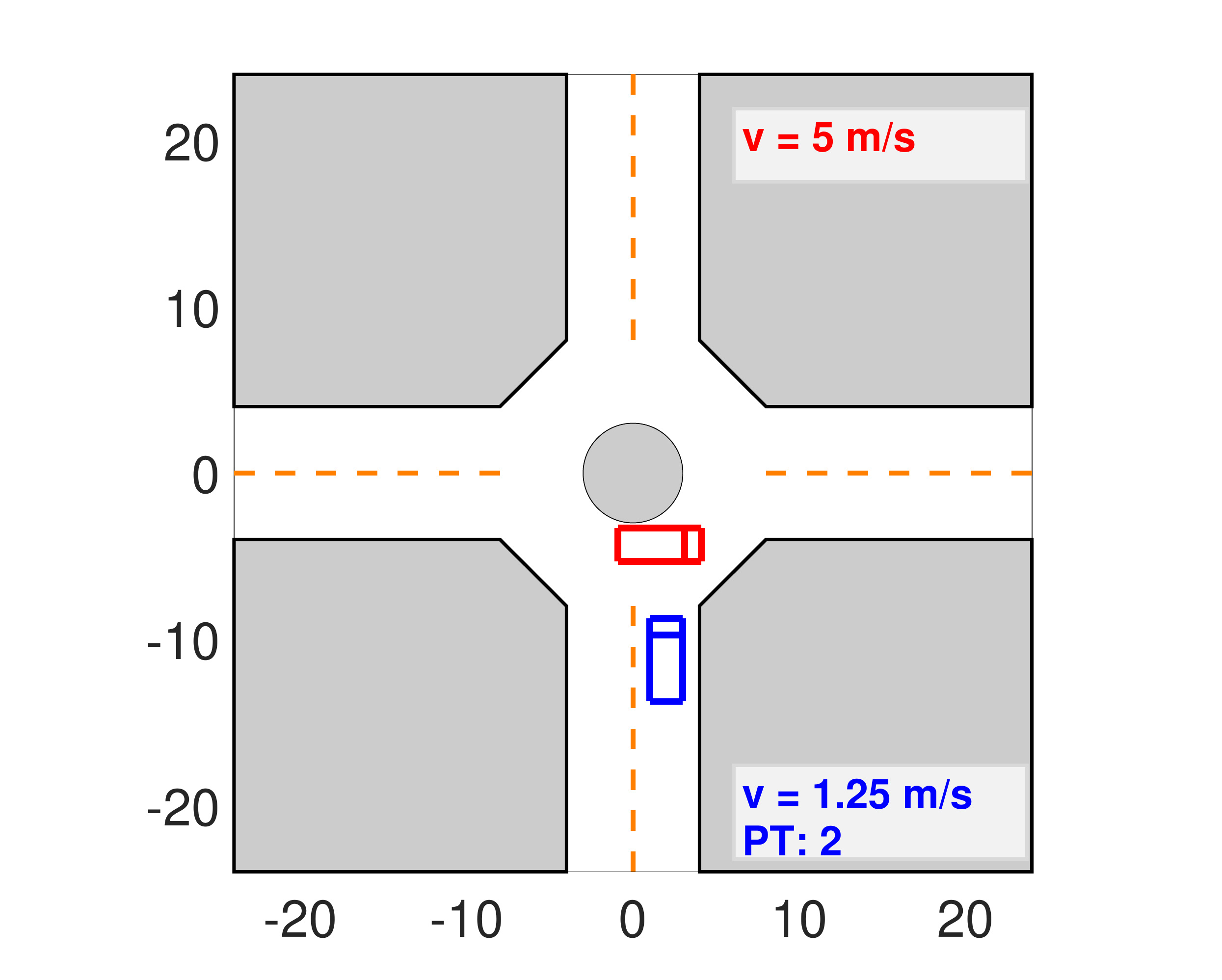,width = 0.4 \linewidth, trim=4.6cm 2cm 0cm 0cm,clip}} 
\put(  10,  94){\epsfig{file=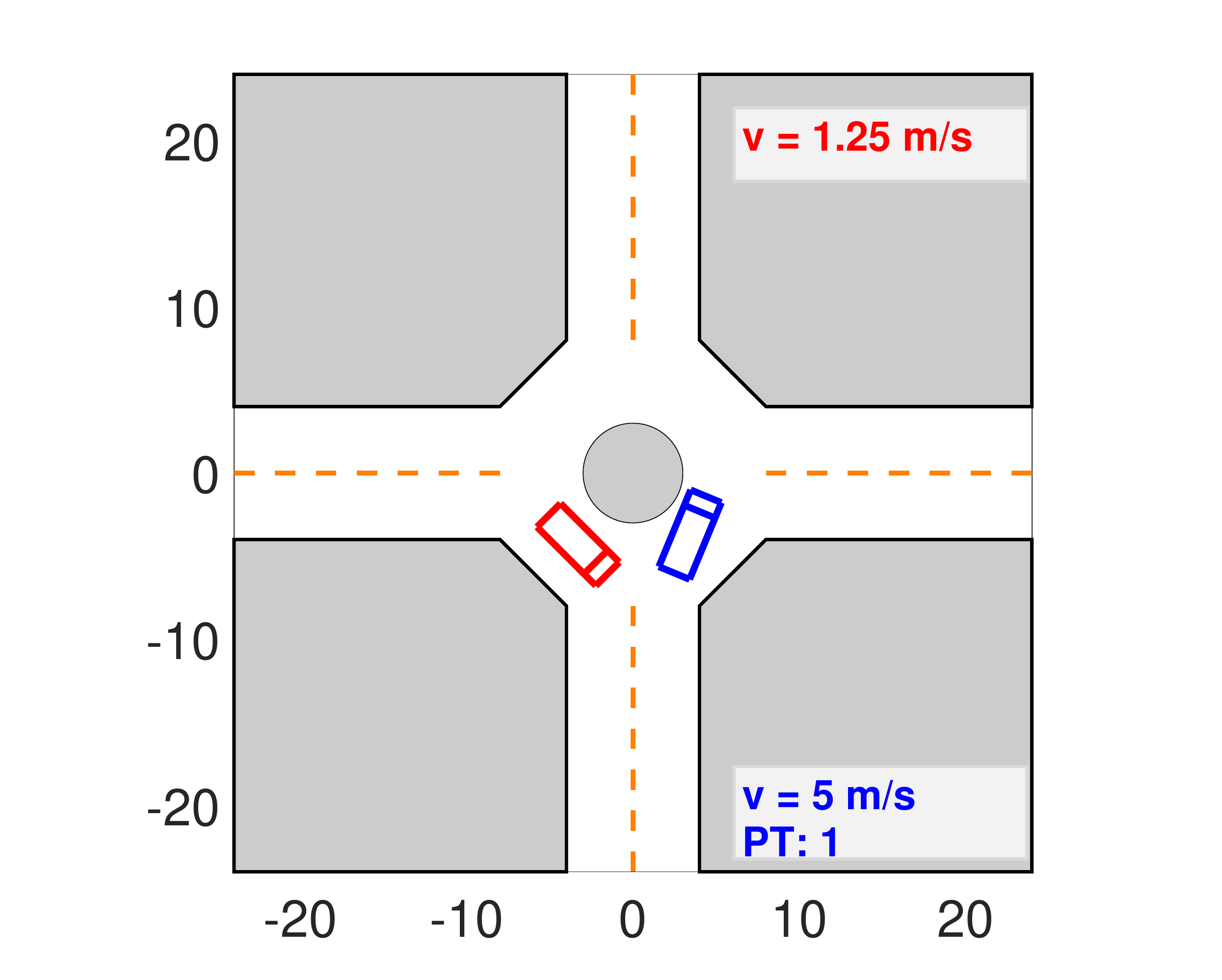,width = 0.4 \linewidth, trim=4.6cm 2cm 0cm 0cm,clip}} 
\put(  120,
94 ){\epsfig{file=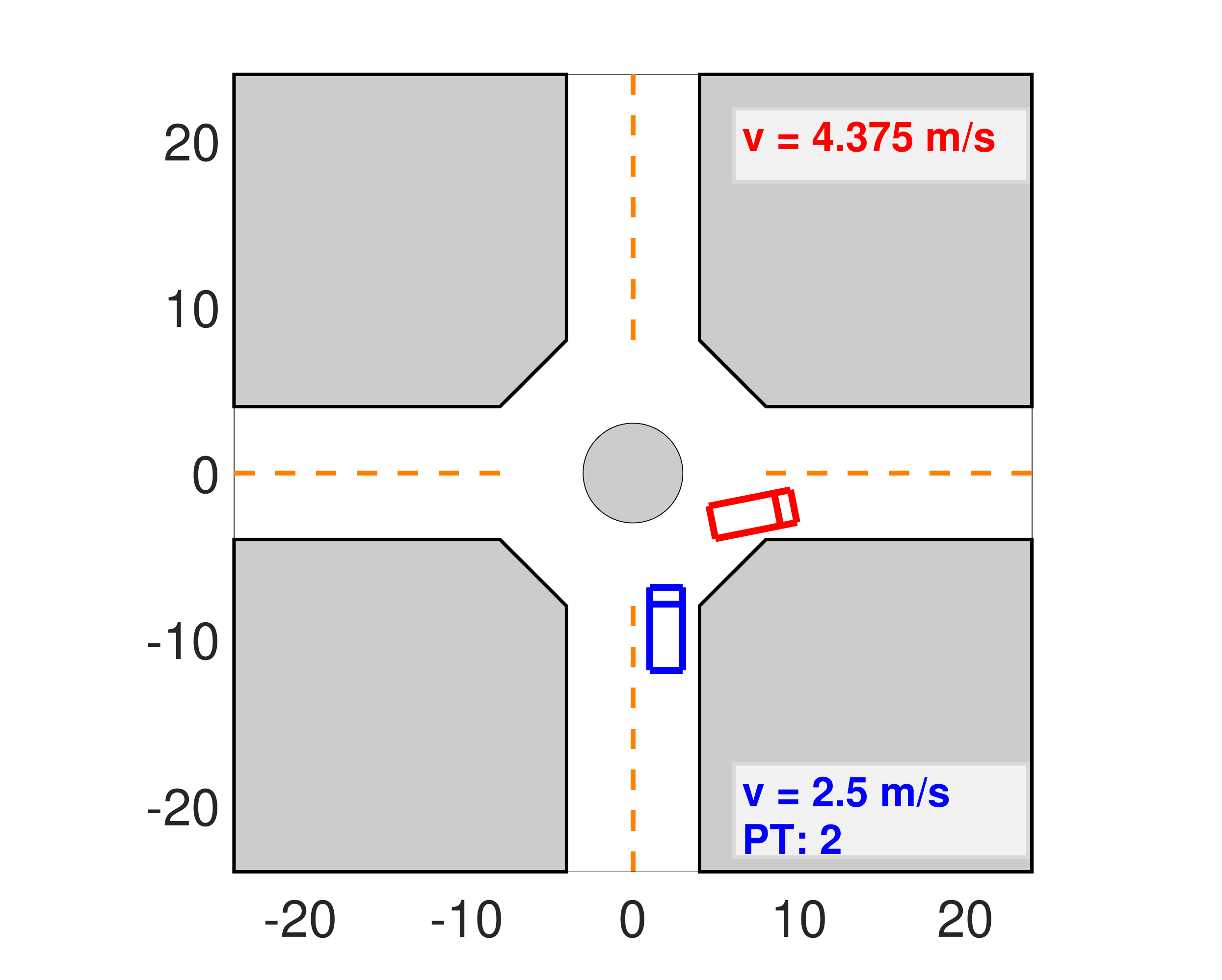,width = 0.4 \linewidth, trim=4.6cm 2cm 0cm 0cm,clip}}
\put(  10,  0){\epsfig{file=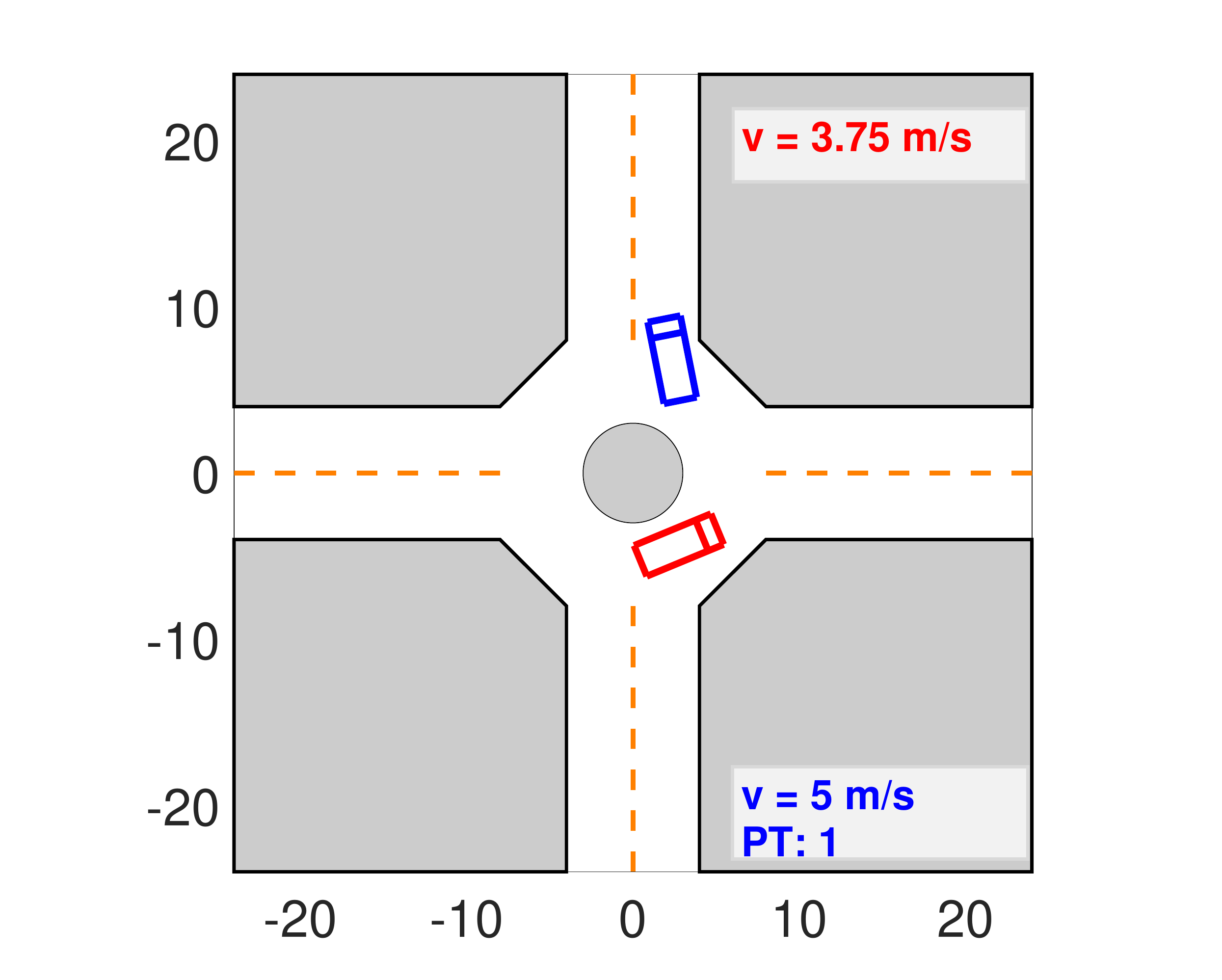,width = 0.4 \linewidth, trim=4.6cm 2cm 0cm 0cm,clip}}  
\put(  120,  0){\epsfig{file=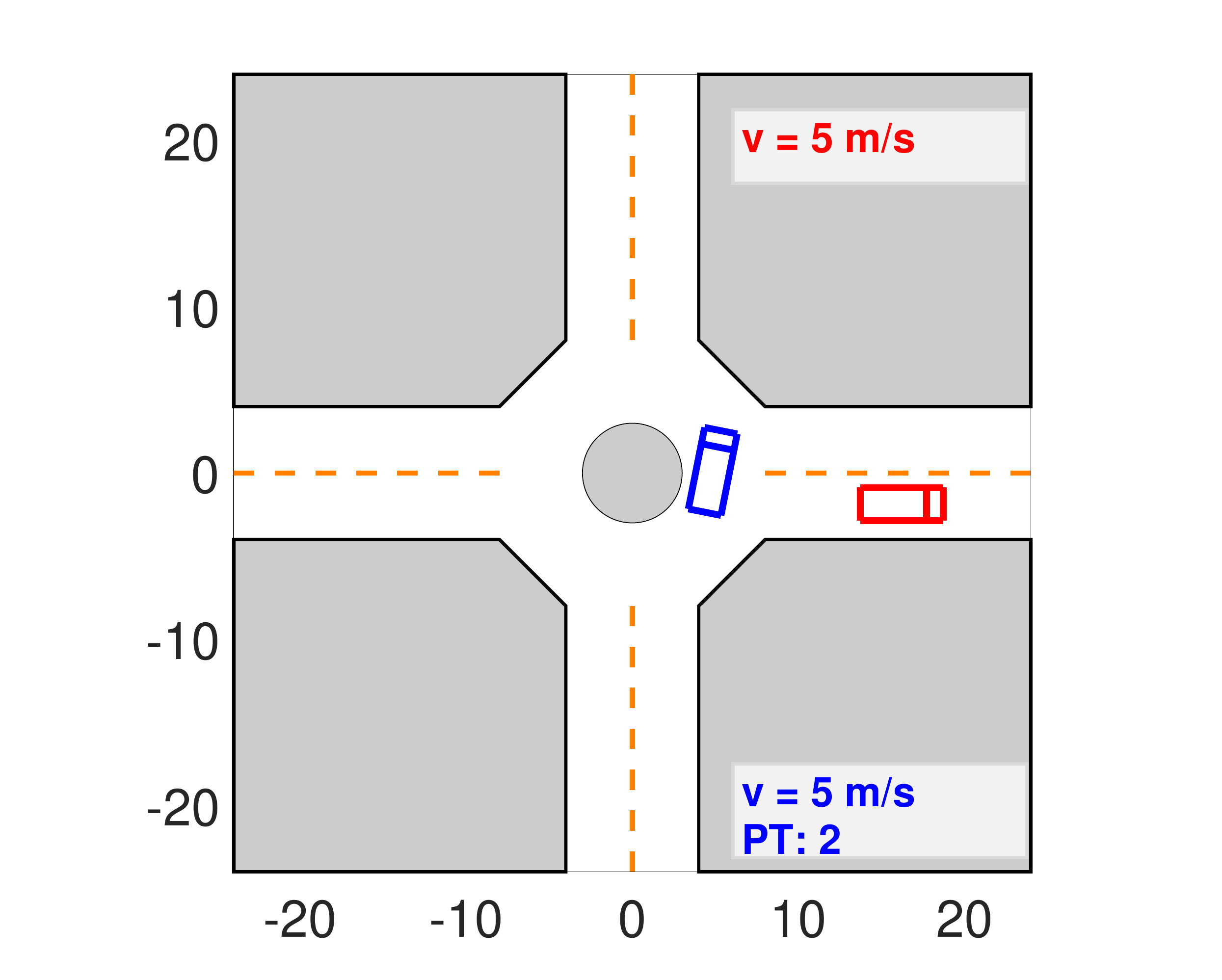,width = 0.4 \linewidth, trim=4.6cm 2cm 0cm 0cm,clip}} 
\small
\put(65,367){(a-1)}
\put(175,367){(b-1)}
\put(65,273){(a-2)}
\put(175,273){(b-2)}
\put(65,179){(a-3)}
\put(175,179){(b-3)}
\put(65,85){(a-4)}
\put(175,85){(b-4)}
\normalsize
\end{picture}
\end{center}
      \caption{Interactions between the 
      ego vehicle (blue) controlled by the proposed AV controller and a type-1 opponent vehicle (red in (a-1) - (a-4)), and a type-2 opponent vehicle (red in (b-1) - (b-4)), at $t = 1.75 \ \text{s}$, $t = 2.5 \ \text{s}$, $t = 3.75 \ \text{s}$ and $t = 6.25 \ \text{s}$.}
      \label{fig: levelk}
\end{figure}
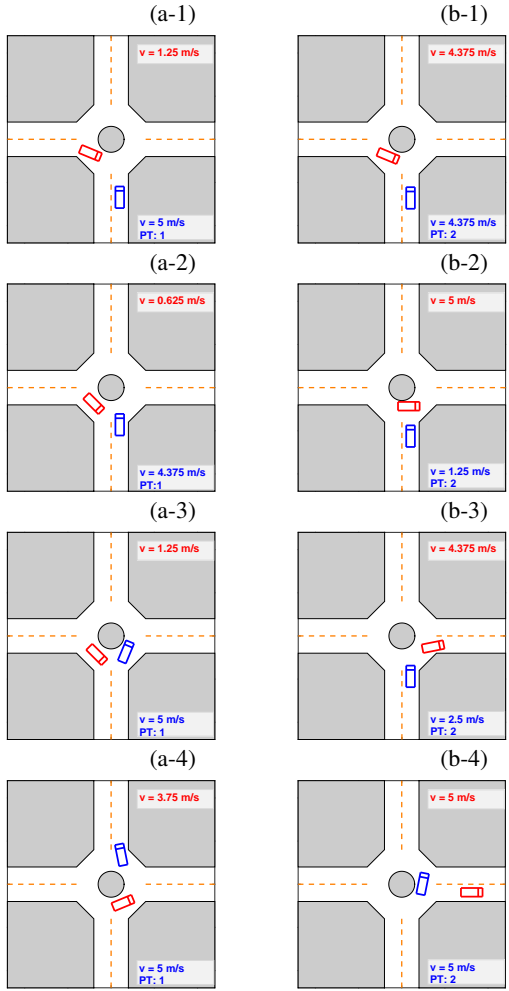

\begin{figure}[!ht]
\begin{center}
\begin{picture}(230, 90.0)
\put(0,-6){\epsfig{file=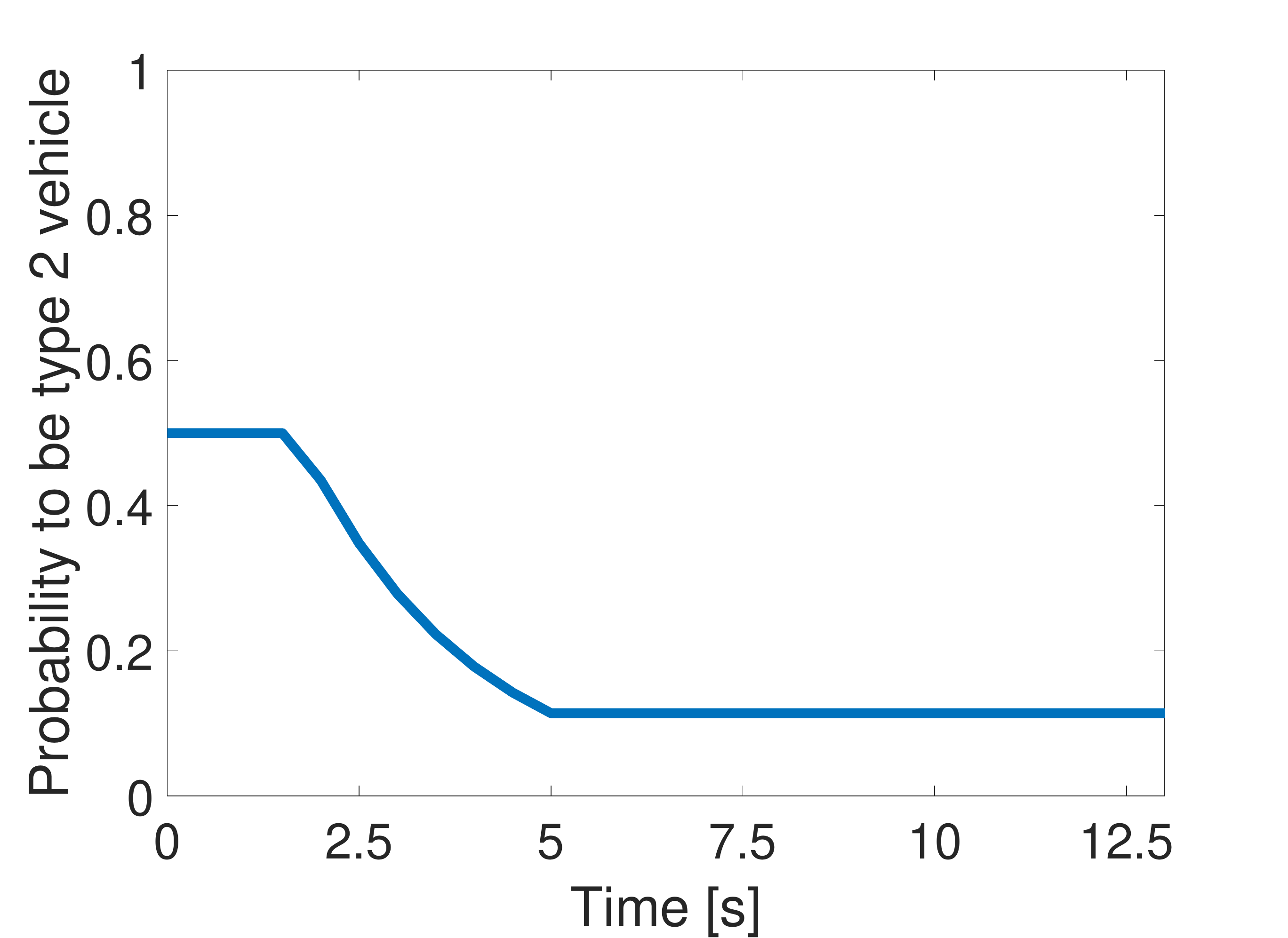, width = 0.5\linewidth}}  
\put(120,-6){\epsfig{file=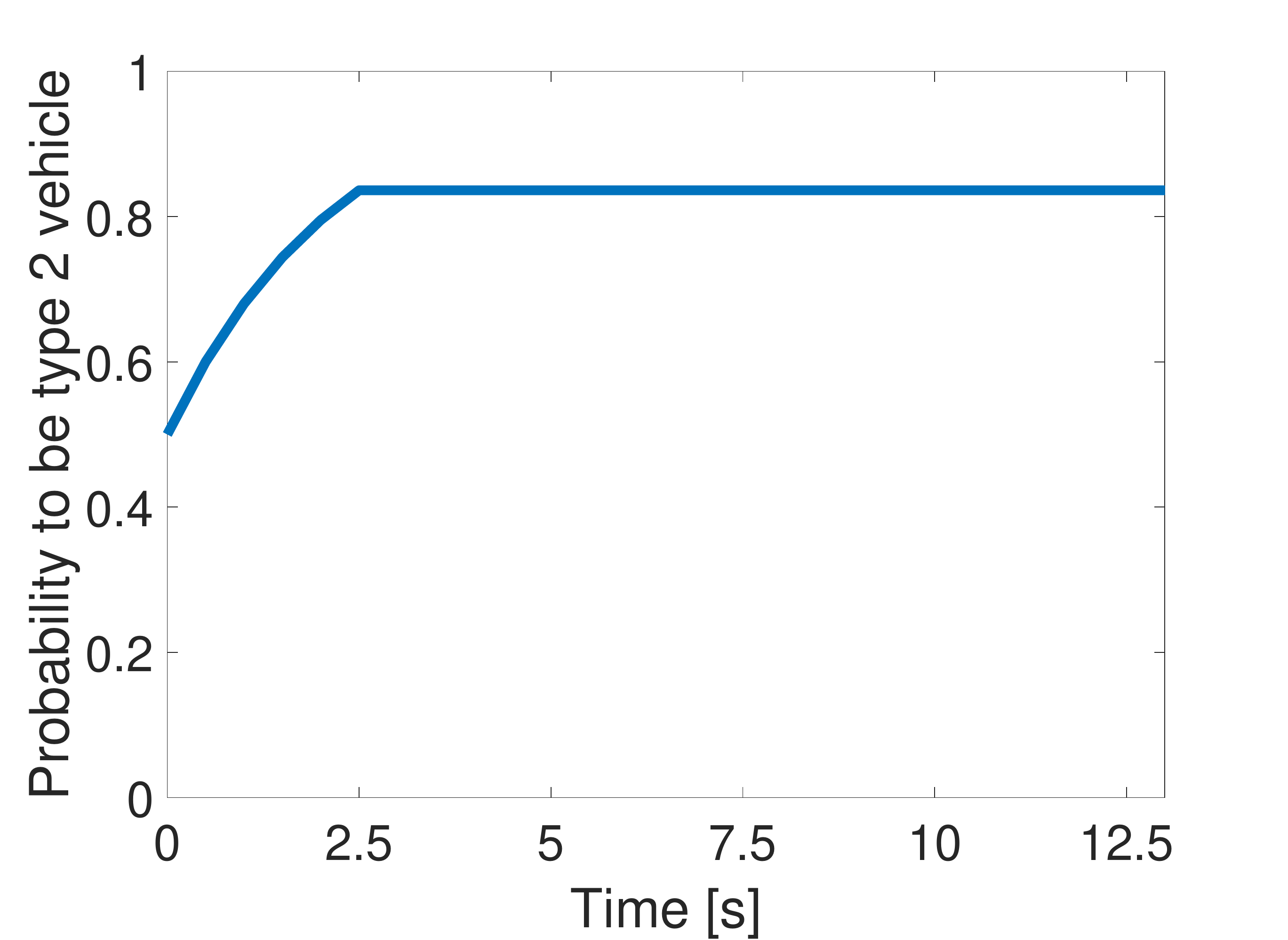, width = 0.5\linewidth}}
\small
\put(95,82){(a)}
\put(215,82){(b)}
\end{picture}
\end{center}
      \caption{Time histories of $\mathbb{P}^{(2)}(t)$ corresponding to the simulations in Fig.~\ref{fig: levelk}. (a) Versus a type-1 opponent vehicle. (b) Versus a type-2 opponent vehicle.}
      \label{fig: L_hist_levelk}
\end{figure}

Fig.~\ref{fig: levelk} (a-1) - (a-4) and (b-1) - (b-4) show the responses of the ego vehicle controlled by the AV controller when it encounters, respectively, a type-1 opponent vehicle and a type-2 opponent vehicle. The initial conditions, $\mathbf{X}(0)$, for both cases are the same. Fig.~\ref{fig: L_hist_levelk} shows the controller's belief histories on the opponent vehicle's type, $\mathbb{P}^{(2)}(t)$, over the simulations. When interacting with a type-1, conservative driver, the ego vehicle chooses to pass the roundabout first, since it predicts that the opponent vehicle will yield the right of way. When interacting with a type-2, aggressive driver, the ego vehicle chooses to decelerate, yields the right of way, and then accelerates to pass the roundabout after the opponent vehicle passes.

We run 1000 simulations, where the initial conditions and the types of the opponent vehicle are randomly generated, to statistically evaluate the controller's performance. The success rate is $93.4\%$, i.e., in 934 out of 1000 simulation runs, the ego and opponent vehicles successfully reach their objective lanes without colliding with each other, without driving off the road or crossing the lane markings that separate traffic of opposite directions, and without causing a deadlock (neither vehicle decides to enter the roundabout or both vehicles get stuck in the middle of the roundabout).

\subsection{AV controller versus human drivers}


We next test the proposed AV controller versus opponent vehicles controlled by human operators. We note that neither do the operators know the mechanism behind the controller, nor does the controller know the operators' driving styles in advance. 



\begin{figure}[!ht]
\begin{center}
\begin{picture}(215.0, 480.0)
\put(  10,  376){\epsfig{file=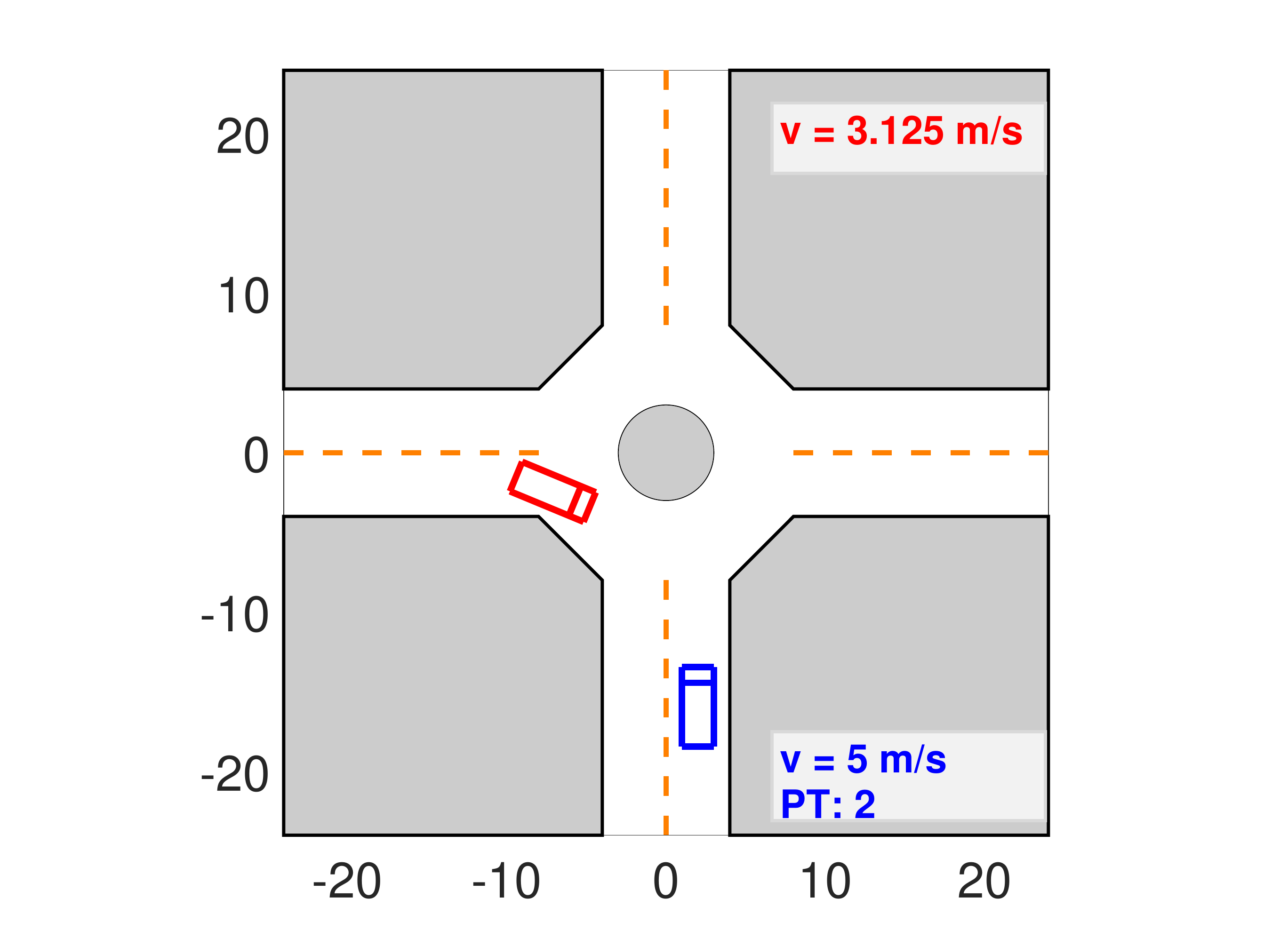,width = 0.425 \linewidth, trim=6cm 2cm 0cm 0cm,clip}} 
\put(  120,  376){\epsfig{file=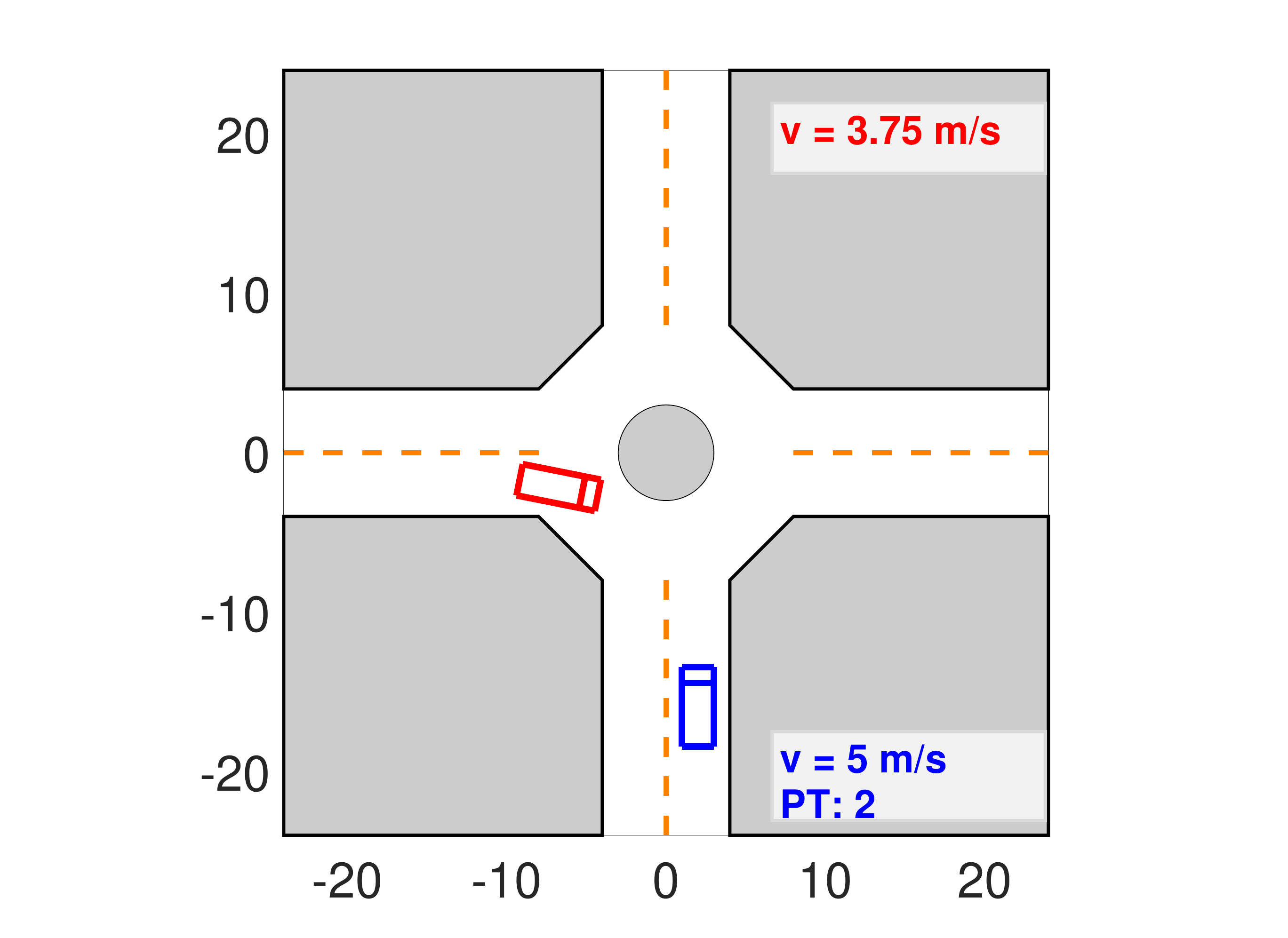,width = 0.425 \linewidth, trim=6cm 2cm 0cm 0cm,clip}}  
\put(  10,  282){\epsfig{file=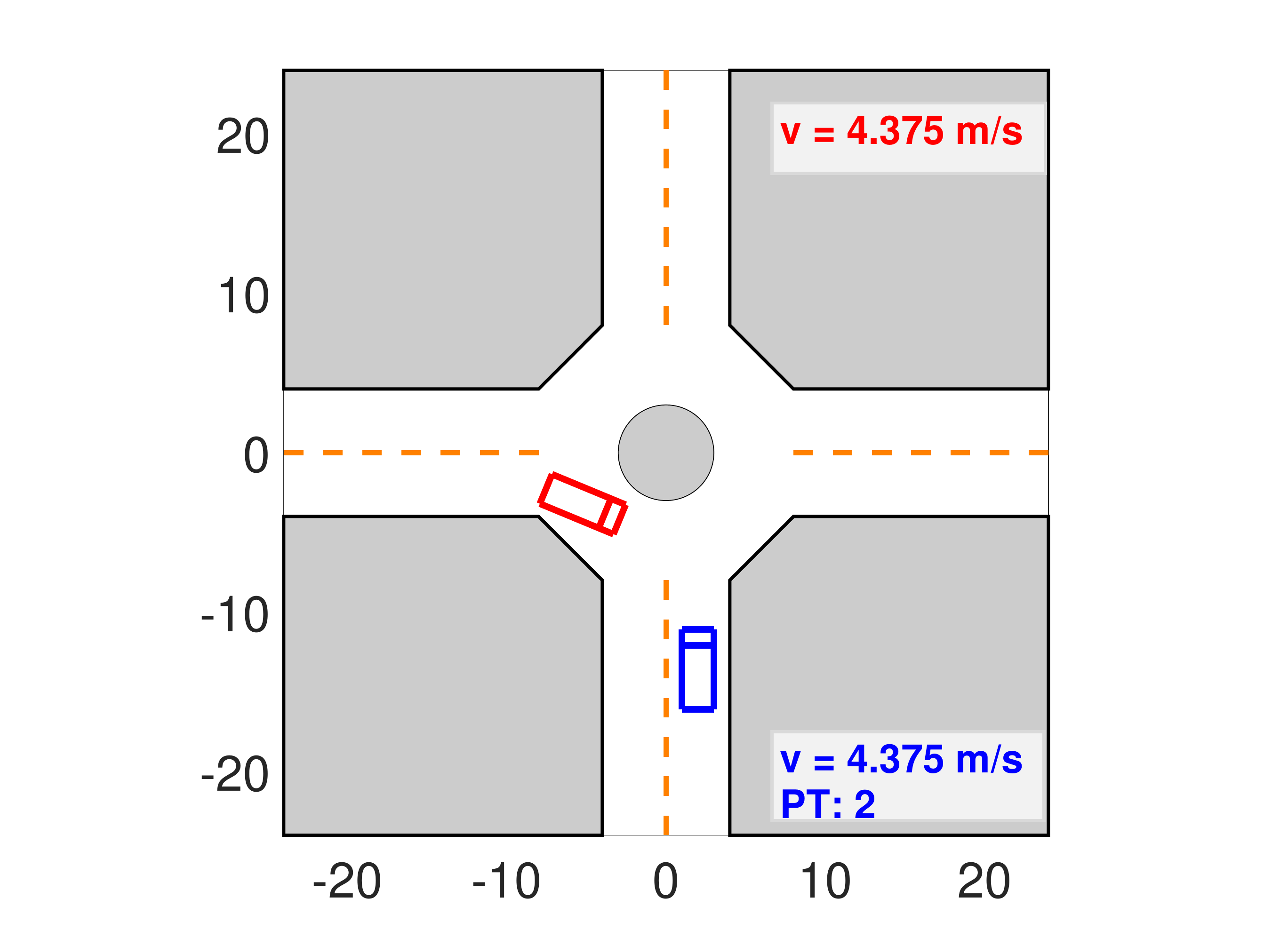,width = 0.425 \linewidth, trim=6cm 2cm 0cm 0cm,clip}} 
\put(  120,  282){\epsfig{file=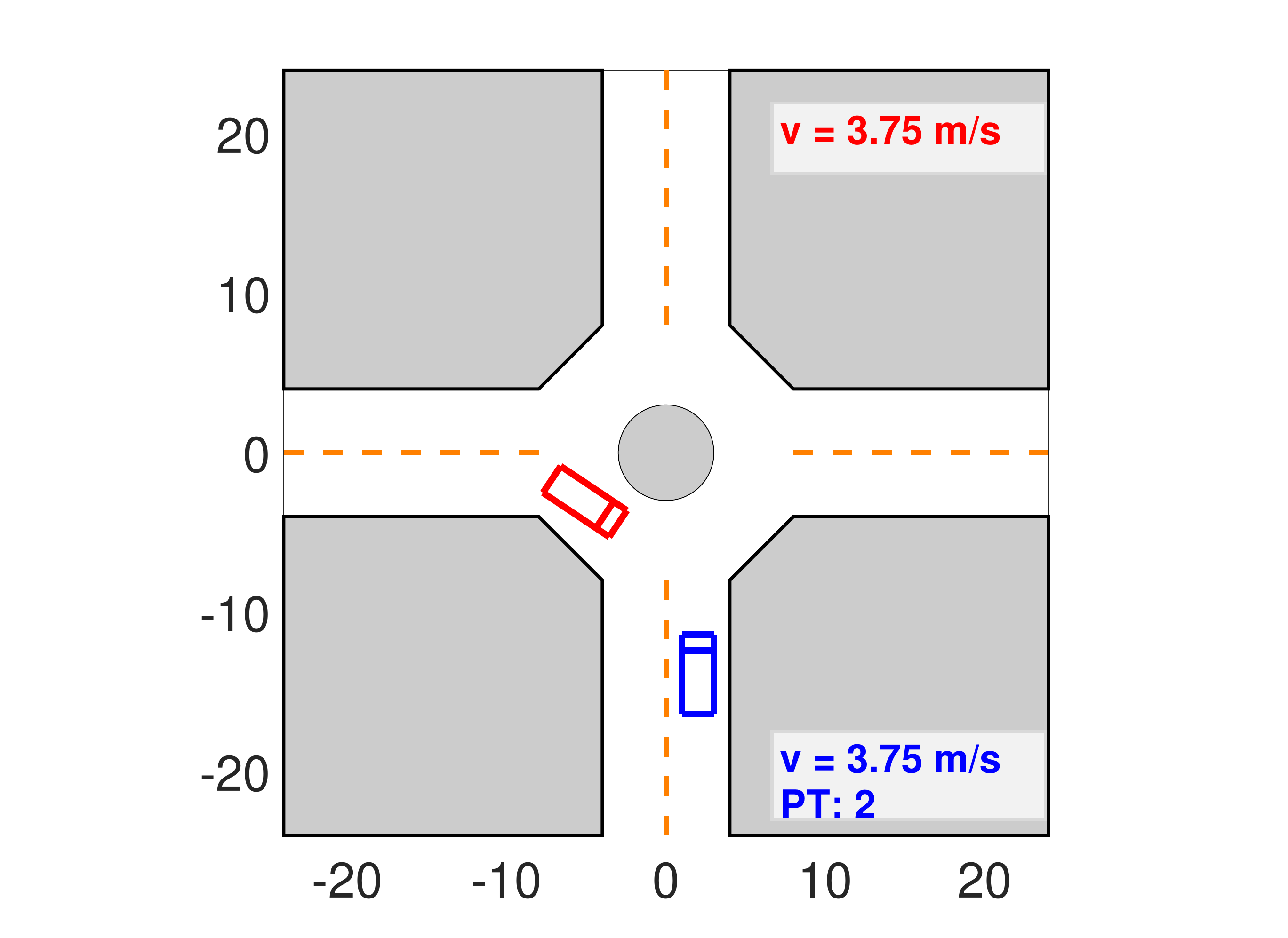,width = 0.425 \linewidth, trim=6cm 2cm 0cm 0cm,clip}}  
\put(  10,  188){\epsfig{file=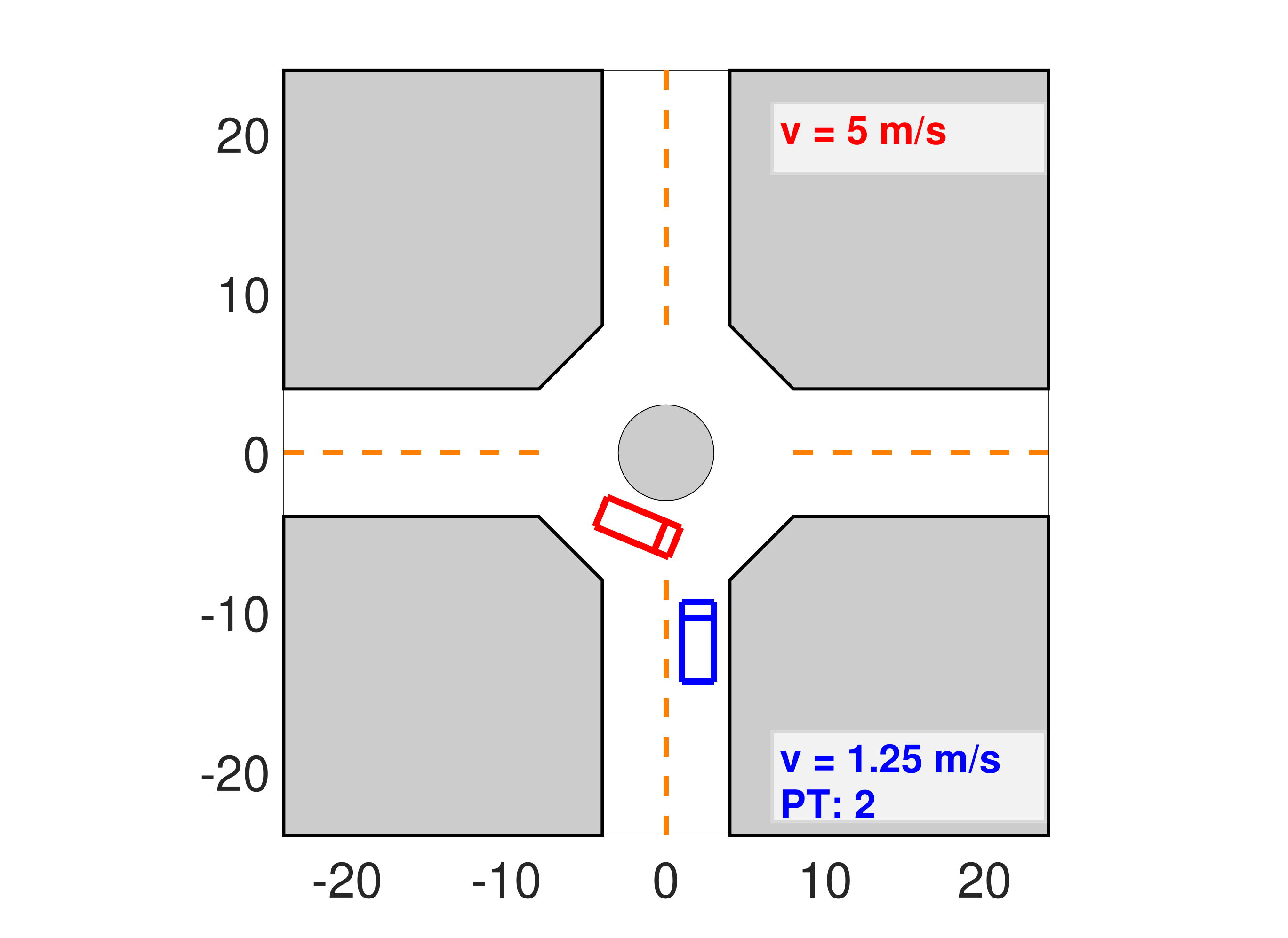,width = 0.425 \linewidth, trim=6cm 2cm 0cm 0cm,clip}}  
\put(  120,  188){\epsfig{file=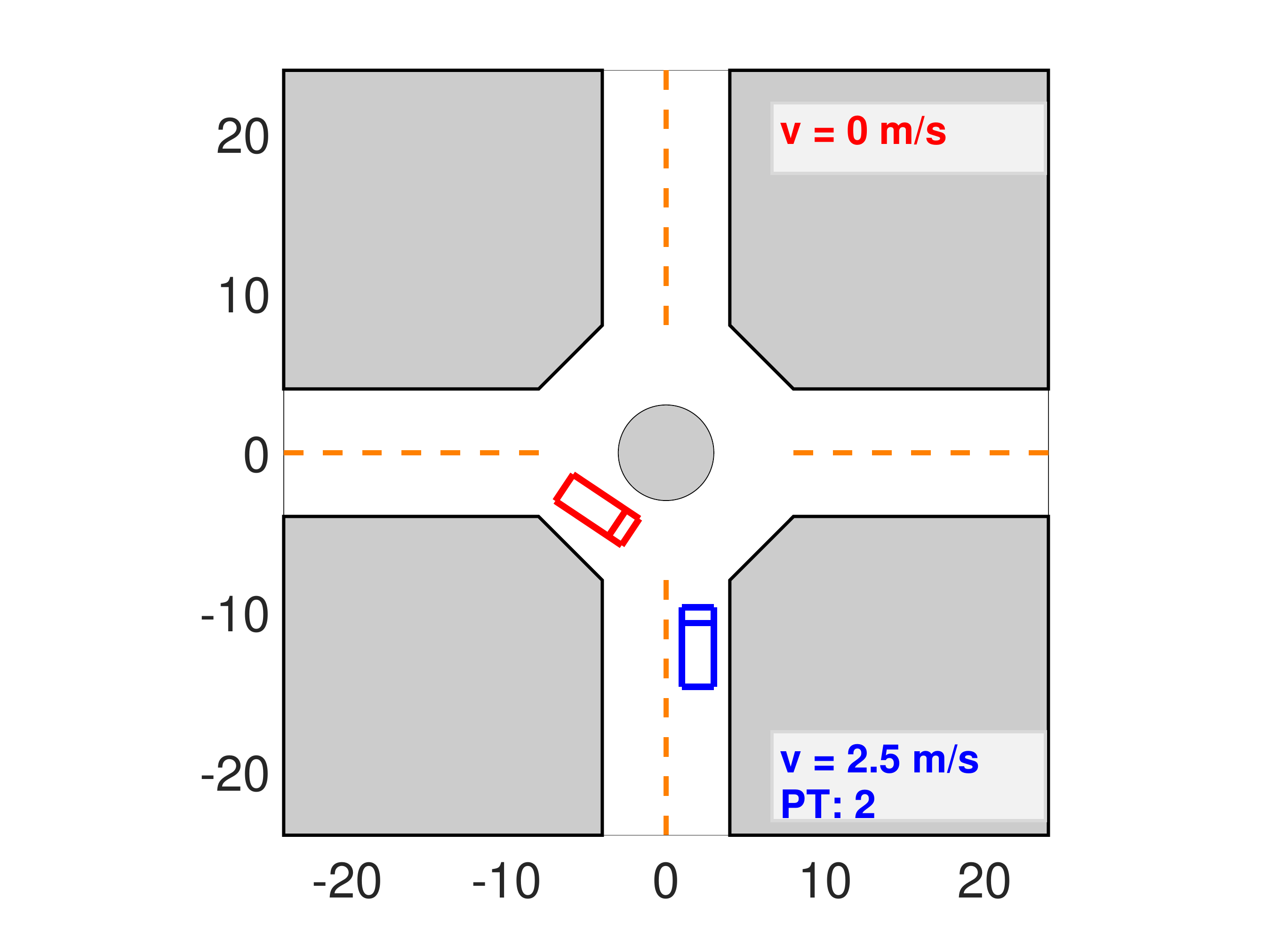,width = 0.425 \linewidth, trim=6cm 2cm 0cm 0cm,clip}} 
\put(  10,  94){\epsfig{file=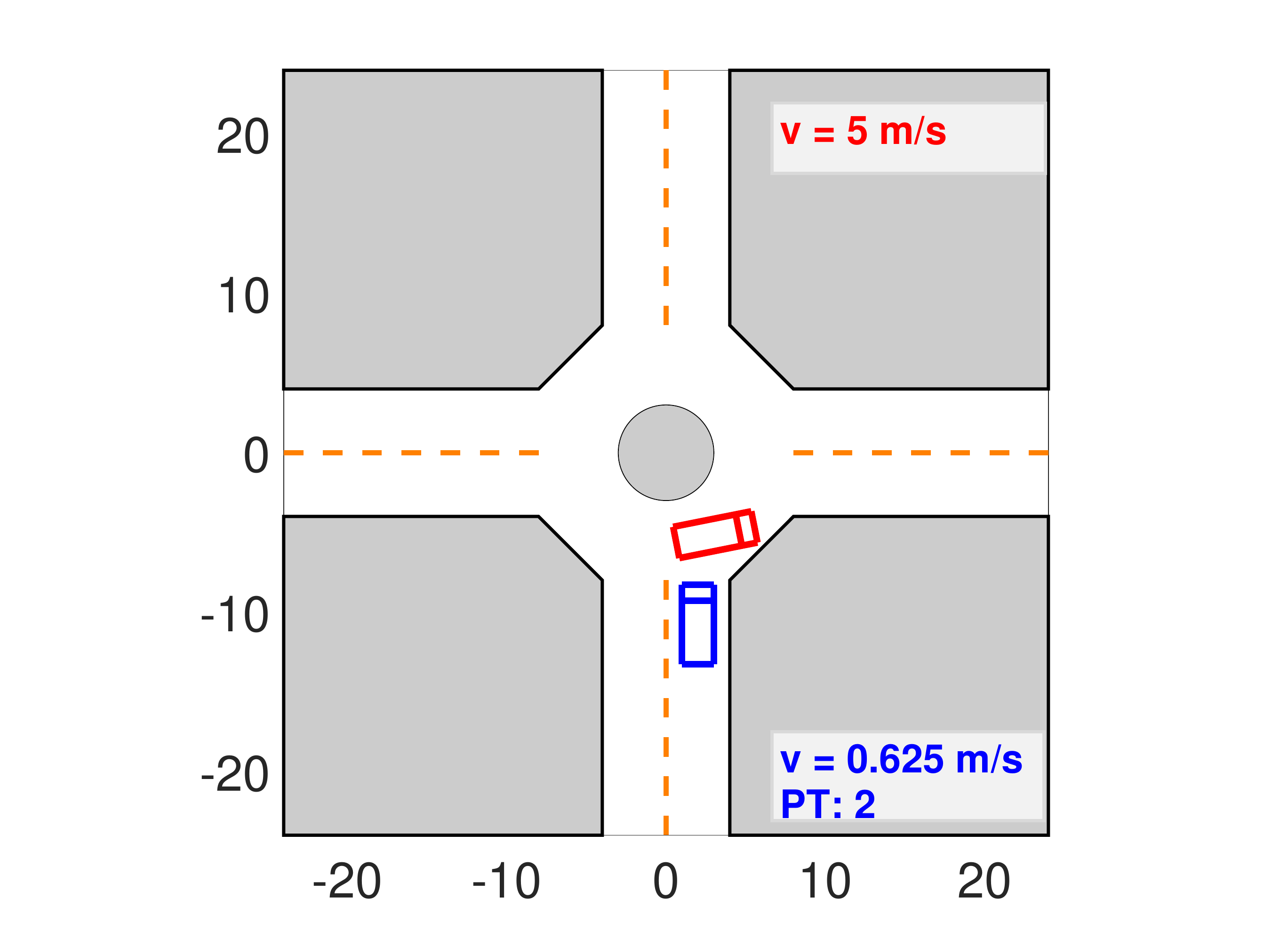,width = 0.425 \linewidth, trim=6cm 2cm 0cm 0cm,clip}}  
\put(  120, 94 ){\epsfig{file=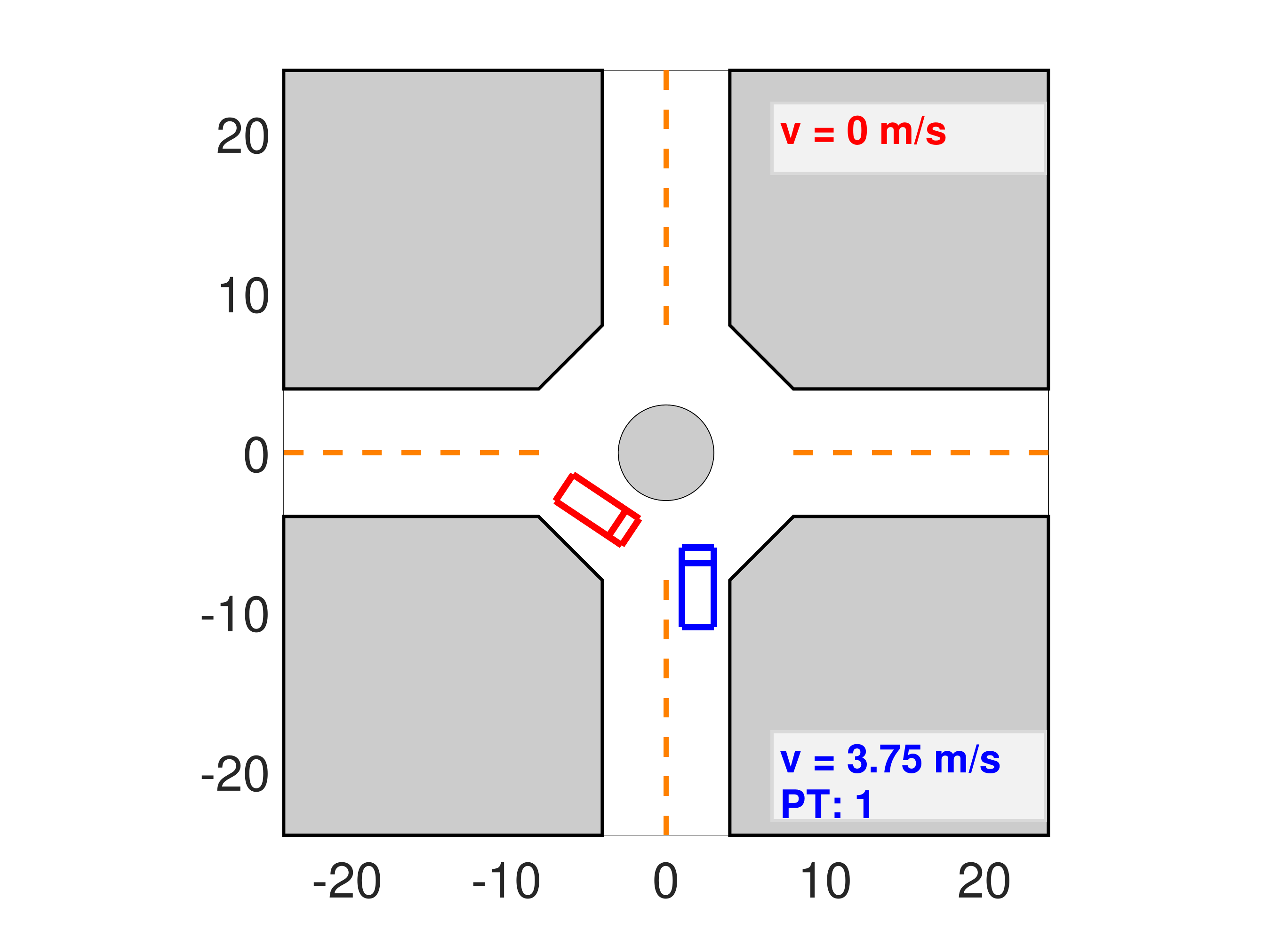,width = 0.425 \linewidth, trim=6cm 2cm 0cm 0cm,clip}} 
\put(  10,  0){\epsfig{file=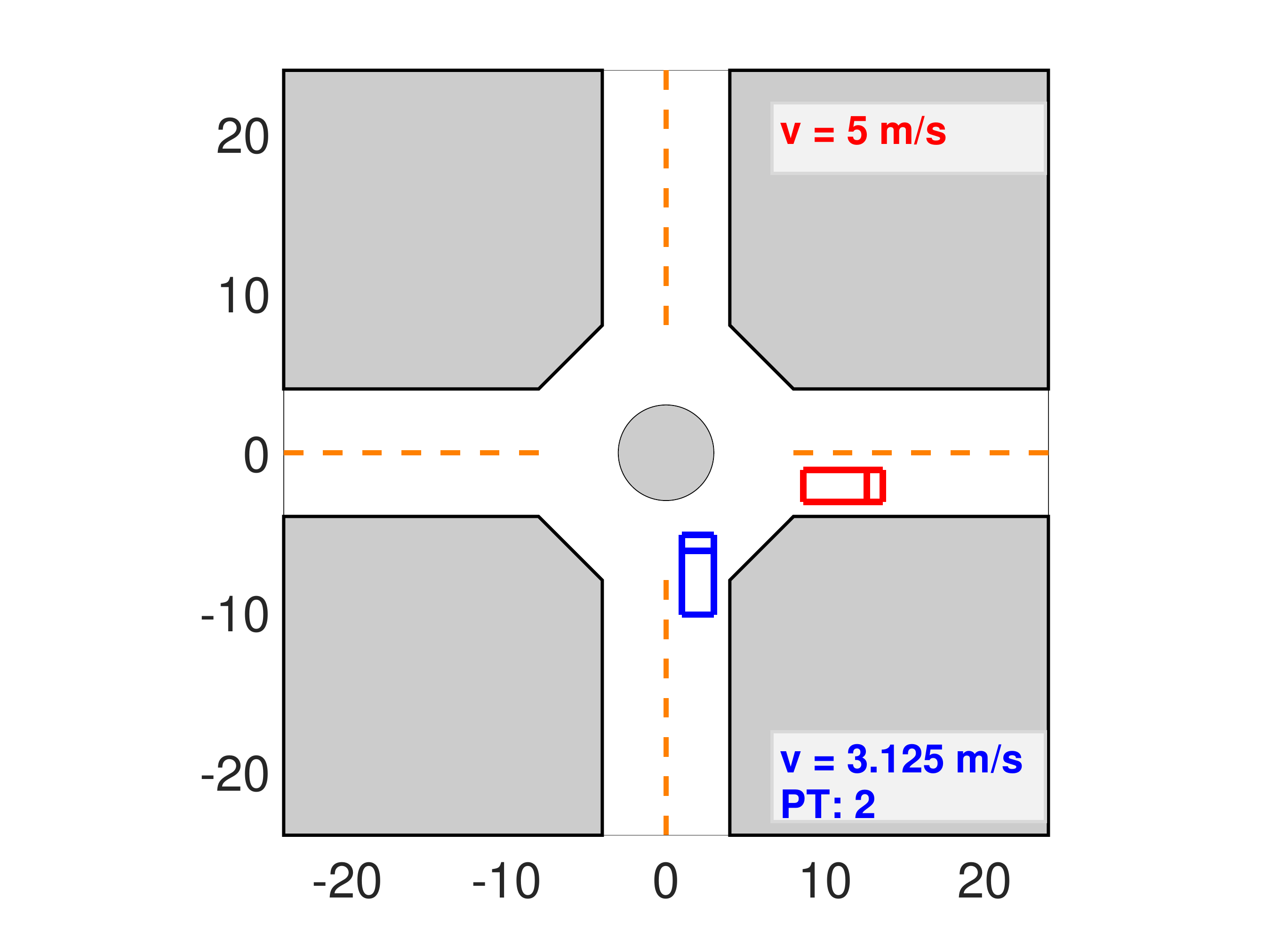,width = 0.425 \linewidth, trim=6cm 2cm 0cm 0cm,clip}}  
\put(  120,  0){\epsfig{file=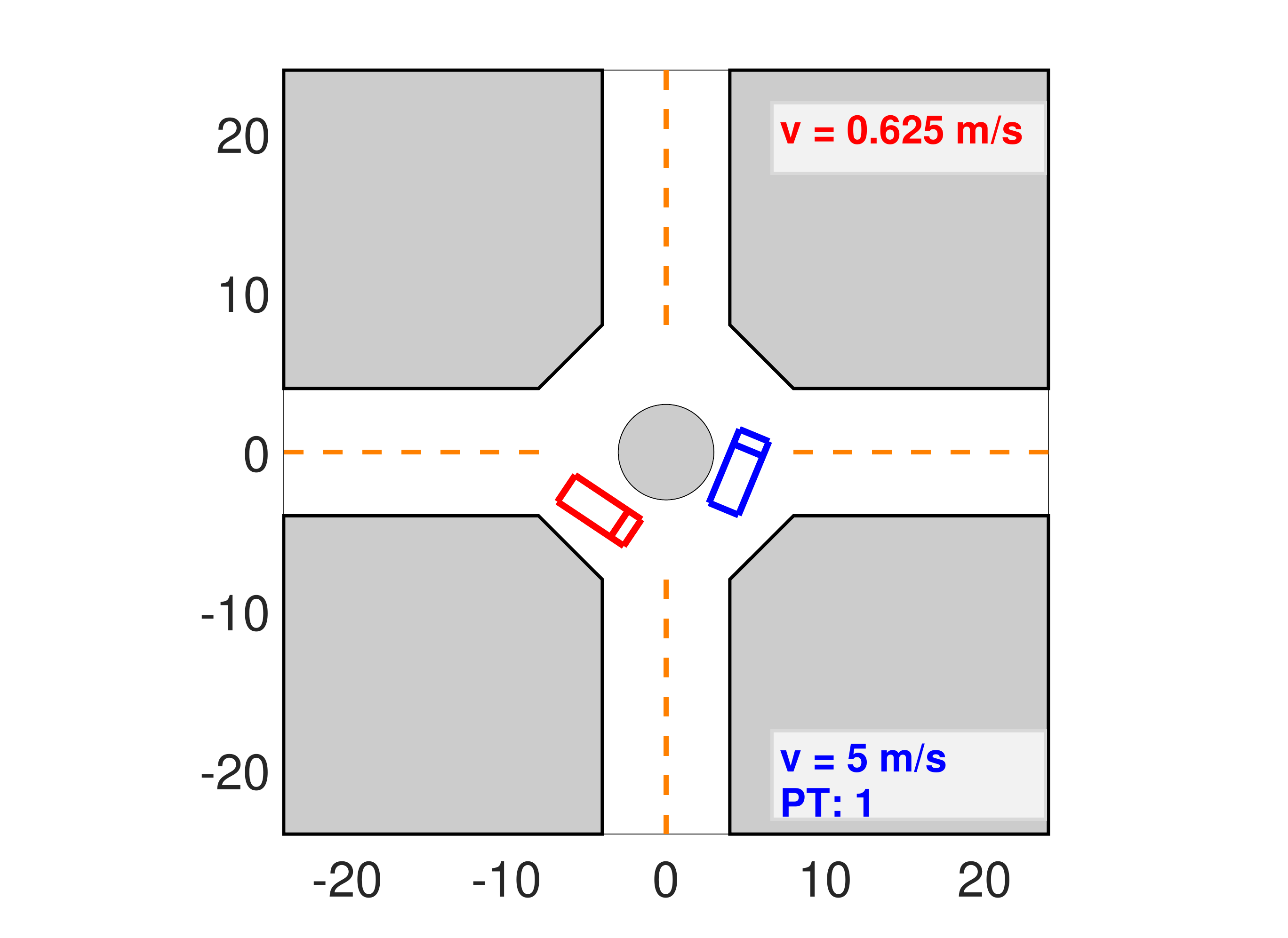,width = 0.425 \linewidth, trim=6cm 2cm 0cm 0cm,clip}} 

\small
\put(65,461){(c-1)}
\put(175,461){(d-1)}
\put(65,367){(c-2)}
\put(175,367){(d-2)}
\put(65,273){(c-3)}
\put(175,273){(d-3)}
\put(65,179){(c-4)}
\put(175,179){(d-4)}
\put(65,85){(c-5)}
\put(175,85){(d-5)}
\normalsize
\end{picture}
\end{center}
      \caption{
      Interactions between 
      the ego vehicle controlled by the proposed AV controller (blue) and opponent vehicles controlled by human operator 1 (red in (c-1) - (c-5)), and by human operator 2 (red in (d-1) - (d-5)), at $t = 1 \ \text{s}$, $t = 1.5 \ \text{s}$, $t = 2.5 \ \text{s}$, $t = 3.25 \ \text{s}$ and $t = 5 \ \text{s}$.}
      \label{fig: driver}
\end{figure}
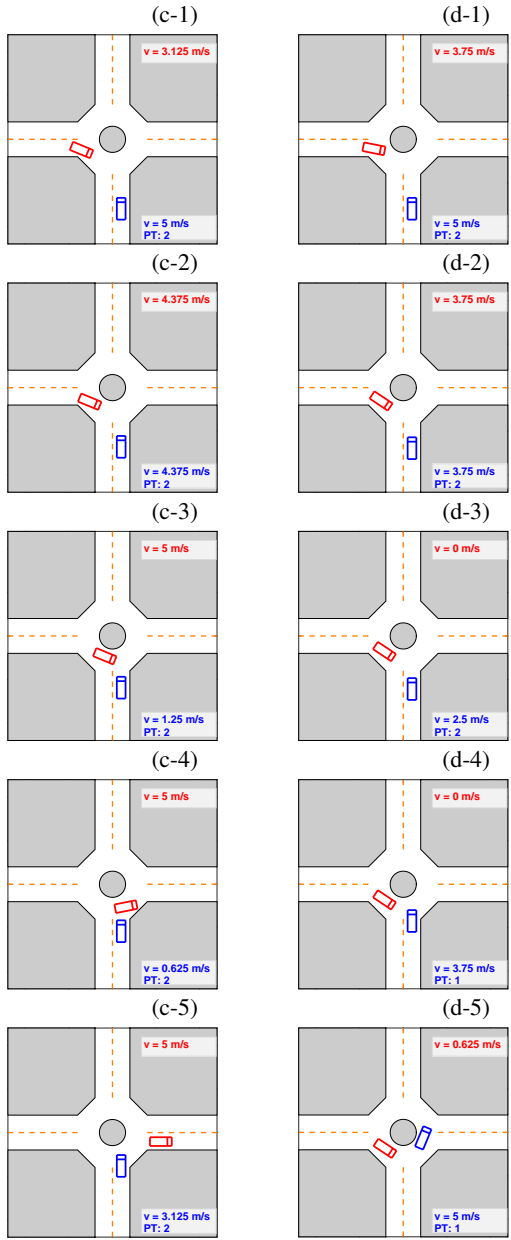

\begin{figure}[ht]
\begin{center}
\begin{picture}(230, 100.0)
\put(0,-6){\epsfig{file=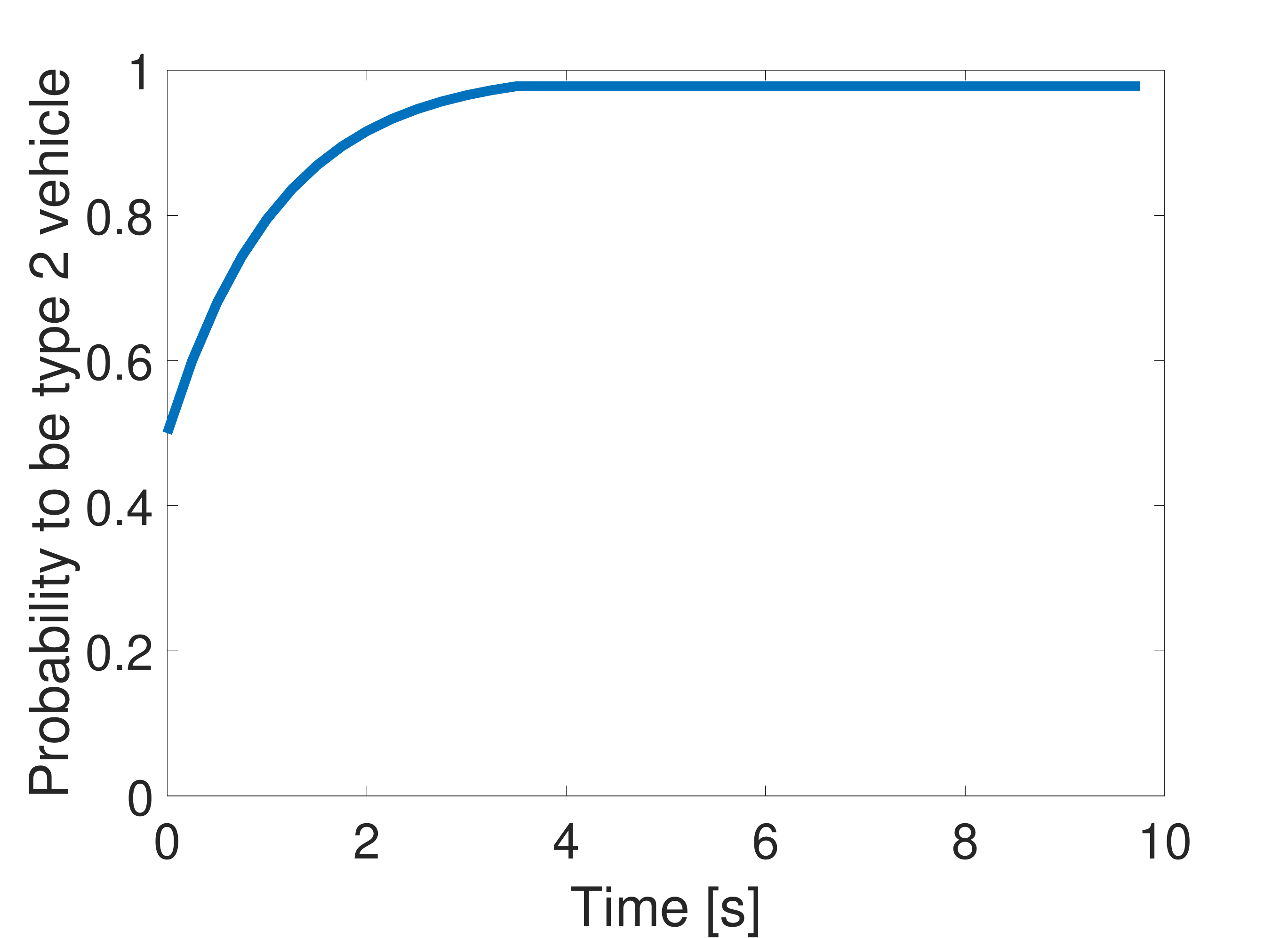, width = 0.5\linewidth}}  
\put(120,-6){\epsfig{file=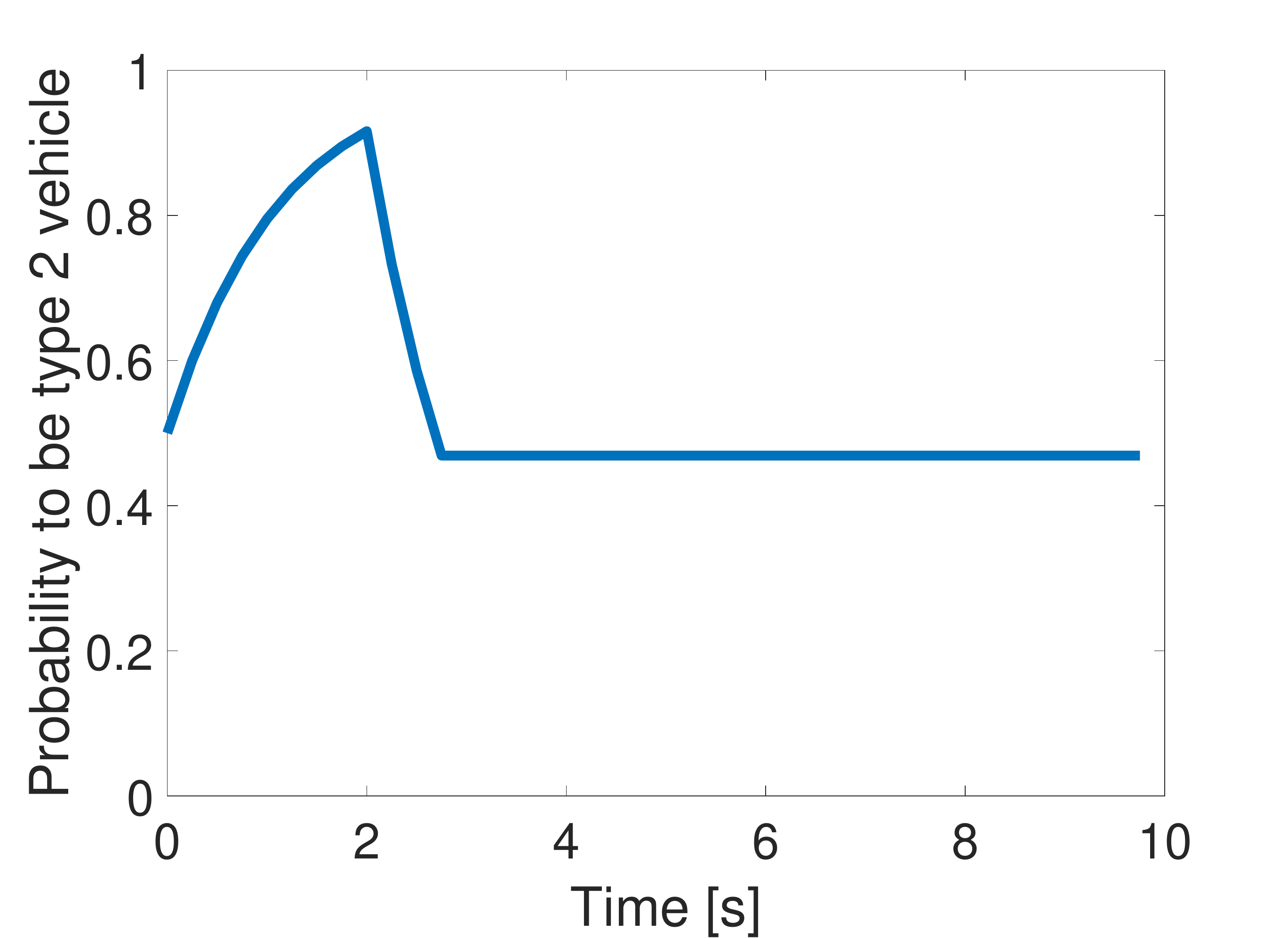, width = 0.5\linewidth}}
\small
\put(95,82){(a)}
\put(215,82){(b)}
\end{picture}
\end{center}
      \caption{Time histories of $\mathbb{P}^{(2)}(t)$ corresponding to the simulations in Fig.~\ref{fig: driver}. (a) Versus human operator 1. (b) Versus human operator 2.}
      \label{fig: L_hist_driver}
\end{figure}

Fig.~\ref{fig: driver} shows the interactions between the ego AV and two opponent human-controlled vehicles, and Fig.~\ref{fig: L_hist_driver} shows the controller's belief histories $\mathbb{P}^{(2)}(t)$ over the simulations. In the first experiment $\big($(c-1) - (c-5)$\big)$, the human operator acts aggressively, so the ego vehicle yields the right of way and the controller identifies the opponent vehicle as type-2 (Fig.~\ref{fig: L_hist_driver}(a)). In the second experiment $\big($(d-1) - (d-5)$\big)$, the human operator accelerates and tries to pass the roundabout first at the beginning. The controller thus identifies the opponent vehicle as type-2 and decelerates to avoid collision. However, the human operator realizes soon that he gets too close to the AV, thus decelerates. Then the AV controller updates its belief on the opponent vehicle's type to type-1 (Fig.~\ref{fig: L_hist_driver}(b)) and accelerates to pass the roundabout first.

We run 70 simulations conducted by 7 different human operators (10 trials for each person) to statistically evaluate the controller's performance. The success rate is $88.6\%$. We note that the human operators may have driven the vehicle more aggressively than they usually do in real driving since there are no safety issues \cite{matthews1998driver}.


We note that our neural network-based online implementation is also computationally feasible -- the average computation time for the AV controller to identify the traffic state status, update the opponent vehicle's type estimate, and then generate the ego vehicle's action, is $34.1 \, \text{[ms]}$ in total, running on a laptop with Intel Core I7 processor and NVIDIA GeForce GTX GPU. 


\section{Concluding Remarks}
\label{sec: summary}

In this paper, we described an algorithm for autonomous vehicle control at a roundabout intersection. The algorithm is based on a game-theoretic model representing the interactions between the ego vehicle and an opponent vehicle, and adapts to an online estimated driver type of the opponent vehicle. We further proposed an explicit online implementation scheme exploiting function approximation techniques. Simulation results were reported to show the feasibility of the proposed control algorithm.

\bibliographystyle{IEEEtran}

\bibliography{ref}

\end{document}